\documentclass[twocolumn,showpacs,preprintnumbers,amsmath,amssymb,prb]{revtex4}

\usepackage{graphicx}
\usepackage{dcolumn}
\usepackage{bm}

\begin{document}


\title{Coarse Grained Density Functional Theories for Metallic Alloys:
Generalized Coherent Potential Approximations and Charge Excess 
Functional Theory.}

\author{Ezio Bruno}
\email{ebruno@unime.it}
\author{Francesco Mammano}
\author{Antonino Fiorino}
\author{Emanuela V. Morabito}

\affiliation{ Dipartimento di Fisica,
Universit{\`{a}} di Messina, Salita Sperone 31, 98166 Messina, Italy.}

\date{\today}

\begin{abstract}
The class of the Generalized Coherent Potential Approximations (GCPA) to the 
Density Functional Theory (DFT) is introduced within the Multiple Scattering
Theory formalism with the aim of dealing
with, ordered or disordered, metallic alloys. All GCPA theories are based on
a common ansatz for the kinetic part of the Hohenberg-Kohn functional
and each theory of the class is specified by an external model concerning
the potential reconstruction. Most existing DFT implementations of CPA based 
theories belong to the GCPA class. The analysis of the formal 
properties of the density functional defined by GCPA theories shows that
it consists of marginally coupled local contributions. Furthermore it is
shown that the GCPA functional
does not depend on the details of the 
charge density and that it can be exactly rewritten as a function of
the appropriate charge multipole moments to be associated with each lattice
site. A general procedure based on the integration of the 'qV' laws 
is described that allows for the explicit construction the same function. 
The coarse grained nature of the GCPA density functional implies a great deal
of computational advantages and is connected with the $O(N)$ scalability
of GCPA algorithms.
Moreover, it is shown that a convenient truncated series expansion of the
GCPA functional leads to the Charge Excess Functional (CEF) theory [E. Bruno, L.
Zingales and Y. Wang, Phys. Rev. Lett. {\bf 91}, 166401 (2003)] which here
is offered in a generalized version that includes multipolar interactions.

CEF and the GCPA numerical results are compared with status of art LAPW 
full-potential density functional calculations for 62, bcc- and fcc-based,
ordered CuZn alloys, in all the range of concentrations. Two facts clearly
emerge from these extensive tests. In first place, the discrepancies between
GCPA and CEF results are always within the numerical accuracy of the
calculations, both for the site charges and the total energies. In second
place, the GCPA (or the CEF) is able to very carefully reproduce the LAPW
site charges and a good agreement is obtained also about the total energies. 

\end{abstract}
\maketitle

\section{Introduction}
After forty years of studies and applications it is now clear that the density functional 
theory (DFT)~\cite{H&K, Mermin} constitutes a formidable tool for the understanding 
of the matter. Nowadays, DFT-based total energy 
calculations~\cite{Dreizler&daProvidencia,Gross&Dreizler,Dreizler&Gross, Gonis} 
and Car-Parrinello molecular dynamics simulations~\cite{CarParrinello} are 
used in a growing number of scientific fields, ranging 
from physics to chemistry to biology. The reason of such an
ubiquitous fortune is that these methods are {\it ab initio},
in the sense that the only underlying models are the fundamental interactions
laws and quantum mechanics. However, just because of their {\it ab initio} nature,
DFT-based methods generally require large computational resources. 
In spite of the availability of faster and faster computers, this circumstance sets up
the limitations to the applicability of the same methods.    
 
Most DFT implementations are based on the Kohn-Sham scheme~\cite{K&S} and
require the solution for the wave-functions of the appropriate Kohn-Sham 
Schroedinger equation. This usually implies the orthogonalization or the 
inversion of large matrices and, hence, a number of operations scaling, 
in principle, as $N^3$, where N is the number of 
atoms in the system. While for semiconductors or insulators, the 
wave-functions localization quite naturally leads to sparse problems, 
the case of metals appears to be the most challenging. 
For metallic systems, in fact, the computational effort required by  
wave-functions based approaches remains $O(N^3)$. 
Nevertheless, even for metals,
approaches based on the direct minimization of the Hohenberg-Kohn functional 
wrt. the charge density~\cite{WeitaoYang,Cortona,Krajewski&Parrinello} could achieve 
$O(N)$ scaling. In this case, the basic strategy consists in partitioning
the system under consideration in a collection of weakly interacting 
fragments~\cite{Harris,Goedecker}. 
Even with nowadays computers, the scaling properties of DFT algorithms wrt. 
the size of the system remain a crucial issue since they determine which 
classes of phenomena can be studied by {\it ab initio} methods.

Among DFT implementations, the oldest methods using such a {\it 'divide and 
conquer'} strategy for metallic systems are perhaps the self-consistent 
versions of the Korringa~\cite{Korringa}, Kohn and 
Rostoker\cite{Kohn&Rostoker} multiple scattering theory (MST). The MST 
method~\cite{Gonis} 
views the system under consideration as a collection of fragments 
(usually in a one to one correspondence to lattice sites) whose 
scattering properties are determined by solving a Kohn-Sham Schroedinger equation. 
Once the fragment (or single-site) scattering matrices are determined, they are 
assembled together with the free electron propagator in order to obtain the 
scattering matrix, or, equivalently, the Green's function of the system. Both 
the determination of the fragment scattering matrices and 
the potential reconstruction are $O(N)$, while, in principle, the solution 
for the global scattering matrix is an $O(N^3)$ problem, as it corresponds 
to the determination of the appropriate boundary conditions for the 
wave-functions in each fragment. However, a number of 
algorithms have been devised~\cite{LSMS,LSGF,swisscheese} 
that are able to obtain $O(N)$ scaling by mapping the determination of 
the system's Green function in a sparse 
problem. This is usually obtained by assuming zero the electronic propagator 
outside the so called Local Interaction Zone (LIZ) of each fragment. If the
free electrons propagator is used, about ten neighbors shells should be included 
in the LIZ~\cite{LSMS}, while using screened propagators~\cite{Lodder}
allows to have much smaller LIZ's: typically one or two neighbors
shells\cite{LSGF} are sufficient. Another remarkable feature of the MST method
is that, being based on Green functions rather than on wave-functions, 
it can deal easily with disordered systems and ensemble statistical averages. 
For this reason, since many years, the Coherent Potential Approximation (CPA) 
theory~\cite{soven} 
for disordered alloys has been used in conjuction with the MST~\cite{KKRCPA} 
and the DFT~\cite{DFTKKRCPA1}.   

The present paper shall be concerned with the study of metallic
alloys in which the nuclei are assumed to occupy the positions of an 
ordered lattice, 
while substitutional disorder may be permitted. For these systems,
in spite of the apparent complexity of the DFT algorithmic implementations, 
the analysis of large supercell calculations has allowed for
the identification of remarkably simple trends. Namely, the charge 
excesses associated with each lattice site appear to be linear functions of
the electrostatic potentials at the same site~\cite{FWS1,FWS2}. These simple relationships,
to be referred in the following to as to the 'qV' laws, allow to describe the 'atoms'
of each chemical species
in an extended metallic system in terms of two parameters, say, the slope and the
intercept of the above linear functions\cite{CPALF,CEF}, and appear to be the appropriate
generalization of Pauling's concept of electronegativity~\cite{Pauling} to 
solid state
physics. We have already suggested that the 'qV' laws can lead to important 
simplifications for total energy calculations in metallic alloys~\cite{CEF}.
In the present paper, we shall introduce the class of the Generalized CPA's (GCPA)
for dealing both with ordered and disordered metallic alloys. From the 
computational point of view, GCPA schemes present $O(N)$ scaling. Their principal
virtue, however, is that, as we shall demonstrate, the GCPA functional {\it exactly}
reduces to a {\it function} of the relevant charge multipole moments at the various
lattice sites, thus constituting a {\it coarse grained} approximate version of
the original DFT. At a further level of approximation, the GCPA density functional
leads to a Ginzburg-Landau functional, the Charge Excesses Functional 
(CEF)~\cite{CEF}, which is
equivalent to the above linear 'qV' laws and computationally inexpensive. 
The predictions of the GCPA and the CEF about the
'qV' laws and total energies shall be compared vs. full-potential Linearized Augmented
Plane Waves (LAPW) calculations~\cite{Andersenlinmet,WIEN2Kb} for  
62 ordered
crystal structures~\cite{Curtarolo,Curtarolothesis}. Our conclusions shall be that,
at least for the systems considered, both GCPA and CEF are generally able to 
find out correctly the system ground state and to fairly well reproduce the 
energy differences between ordered structures in a fixed concentration 
ensemble.        

The following of this paper is organized as follows. In Sect. II we shall
briefly review the MST version of the DFT. In order to have a 
functional form as much localized as possible, the relevant electrostratic 
contributions shall be rewritten using an exact multipole expansion.
In Sect. III, we shall introduce the class of the 
GCPA theories and investigate the analytical properties of the corresponding
approximate density functional. Moreover we shall obtain, as a further
approximation to the GCPA, the CEF
theory, already obtained in a much more phenomenological
context~\cite{CEF}, that here is offered in a generalized form suitable for the inclusion
of dipole or quadrupole interactions. In Sect. IV we shall compare the
numerical results obtained from the GCPA and CEF approximations with
those from full-potential LAPW calculations. CEF and GCPA calculations appear 
{\it numerically indistinguishable} one from the 
other and both theories appear able to fairly well reproduce the LAPW total energies.
In the final Sect. V, we shall draw our conclusions, make our comments
and briefly discuss the possible developments of CEF and GCPA theories.

\section{Review of The Density Functional Multiple Scattering Theory}
\label{SectII}
\subsection{The MST formalism}
\label{SectIIA}
In this subsection we shall briefly overview the grand 
canonical ensemble formulation of the MST-DFT~\cite{Mermin,Dreizler&Gross}. 
Our aim shall be developing a common ground for dealing both with ordered and 
substitutionally
disordered systems. Although finite temperature, relativistic and magnetic 
generalizations could straightforwardly be carried out~\cite{Dreizler&Gross}, 
in this paper we focus on the non-relativistic, non 
spin-polarized case at $T=0$. Furthermore, when not otherwise stated, we 
shall consider the Local Density Approximation (LDA)~\cite{Dreizler&Gross} to the DFT
and assume to have ions of charge $+eZ_i$ fixed at the lattice 
positions $\mathbf{R}_i$.

In our discussion, the relevant density functional is the electronic grand potential~\cite{
DFTKKRCPA1,DFTKKRCPA2},
\begin{eqnarray}
\label{gpotNmu}
&\Omega&(T=0,V,\mu)=E_{TOT}-\mu N(\mu)= \nonumber \\
&-&\int_{-\infty}^{\mu} d \varepsilon \; N(\varepsilon;\mu) 
+\int_{-\infty}^{\mu} d \mu^\prime \int_{-\infty}^{\mu^\prime} d \varepsilon \; 
\frac{d N(\varepsilon;\mu^\prime)}{d \mu^\prime} \nonumber \\
&+&\frac{e^2}{2} \sum_{i,j \; (i \ne j)} \frac{Z_i Z_j}{R_{ij}} 
\end{eqnarray}
where $V$ is the volume of the system, $\mu$ is the chemical 
potential and $E_{TOT}$ is the sum of the total electronic energy and 
the nuclei electrostatic interaction. $N(\varepsilon;\mu)$ is
the integrated density of states which is related to 
to the electronic density of states (DOS), $n(\varepsilon,\mu)$, through the following
relationship: 
\begin{equation}
\label{NmuT0}
N(\varepsilon,\mu)=\int_{-\infty}^{\varepsilon} d \varepsilon \; n(\varepsilon,\mu) 
\end{equation}
The notation highlights the implicit $\mu$ dependence of the DOS that arises  
from the effective Kohn-Sham potential. In a frozen ions treatment, of course,
the nuclear interactions term is just a constant that is included here for 
future convenience. 

The basic idea underlying the MST is partitioning the system in 'small' 
scattering volumes, $v_i$, $\sum_i v_i=V$, which in most 
implementations are 'centred' at the nuclei positions. Although at this
stage the partitioning is quite arbitrary, as we shall see in the following,
there is a natural choice for it.
Using the Lloyd's formula~\cite{Lloyd,FaulknerLloyd}, the integrated DOS, 
$N(\varepsilon;\mu)$, 
can be expressed as the excess with respect to the corresponding
free electrons quantity, $N^0(\varepsilon)$: 
\begin{eqnarray}
\label{Lloyd1}
&&N(\varepsilon;\mu)=N^0(\varepsilon)-\frac{1}{\pi} Im \ln \det 
\underline{\underline{M}}(\varepsilon)= \nonumber \\
&&N^0(\varepsilon)+\frac{1}{\pi} Im \sum_i Tr (\ln \underline{\underline{\tau}}(\varepsilon))_{ii} 
\end{eqnarray} 
where the trace is taken only over the angular momentum components.
In Eq.~(\ref{Lloyd1}) the multiple scattering matrix,
$\underline{\underline{M}}$, or the scattering-path 
matrix~\cite{Gyorffy&Stott},
$\underline{\underline{\tau}}=\underline{\underline{M}}^{-1}$ 
are defined in terms of the 
single-site scattering matrices \footnote{Matrices in the angular 
momentum space are denoted by a single underline, 
double underline is used for matrices both in the angular
momentum and in the site spaces. Angular momentum components are 
denoted by capital letters, $L \equiv (\ell,m)$, 
$L^\prime \equiv(\ell^\prime,m^\prime)$, 
$\dots$, site components by small Latin letters, 
$i, j, \dots$.}, $\underline{t}_i(\varepsilon)$, and the 
free electron propagator, $\underline{G}_{ij}^0(\varepsilon)$, is given by:
\begin{equation}
\label{MKKR}
\underline{M}_{ij}(\varepsilon)=\underline{t}_i^{-1}(\varepsilon) \delta_{ij}-
\underline{G}^0_{ij}(\varepsilon) 
\end{equation} 
It is convenient to recall here that the single-site scattering matrices
convey the informations about the phase shifts at the surfaces 
delimiting each scattering volume. The continuity of the wave-functions 
at the same surfaces is ensured by the construction of the 
scattering-path matrix, $\underline{\underline{\tau}}$, this is
accomplished by the numerical inversion of the multiple 
scattering matrix $\underline{\underline{M}}$. Since the size of 
$\underline{\underline{M}}$ is proportional to the number of scatterers
in the problem, its inversion is the 
source of $O(N^3)$ scaling in the MST version of the DFT.

Within MST the link between the electronic density and the
scattering matrices is provided by the Green function~\cite{Gonis} 
\begin{eqnarray}
\label{greenfun}
 &&G_{ij,LL^\prime}(\mathbf{r},\mathbf{r}^\prime,\varepsilon)=
Z_{i,L}(\mathbf{r},\varepsilon) \tau_{ij,LL^\prime}(\varepsilon)  
Z_{j,L^\prime}(\mathbf{r}^\prime,\varepsilon) \nonumber \\
&&- \Big[ \theta(r-r^\prime) Z_{i,L}(\mathbf{r},\varepsilon) 
J_{j,L^\prime}(\mathbf{r}^\prime,\varepsilon) \nonumber \\
&&+\theta(r^\prime-r) J_{i,L}(\mathbf{r},\varepsilon) 
Z_{j,L^\prime}(\mathbf{r}^\prime,\varepsilon)
\Big] \delta_{L L^\prime}
\delta_{i j}  
\end{eqnarray}
where $\mathbf{r} \; \epsilon \; v_i$, $\mathbf{r}^\prime \; \epsilon \; v_j$. 
$Z_{i,L}(\mathbf{r},\varepsilon)$ 
and $J_{i,L}(\mathbf{r},\varepsilon)$ are, respectively, 
the regular and irregular at $\mathbf{r}=0$
solutions\footnote{In the present paper the convention is used that for any 
function of the position $f_i(\mathbf{r})$ stands for $f(\mathbf{r-R}_i)$ 
with $\mathbf{r} \; \epsilon \; v_i$.} 
of the KS Schroedinger equation for the energy $\varepsilon$. For real 
energies both 
$Z_{i,L}(\mathbf{r},\varepsilon)$ and $J_{i,L}(\mathbf{r},\varepsilon)$ are
real functions. The (site resolved) charge densities, the 
DOS and the neat charges at the $i$-th site can be obtained 
by integrating the Green function over the energy and/or the appropriate volumes 
and by taking the trace over the angular momentum indexes as follows: 
\begin{equation}
\label{rhoi}
\rho_i(\mathbf{r} ; \mu)=-\frac{1}{\pi} \int_{-\infty}^{\mu} d \varepsilon \sum_L 
Im \{ G_{ii,LL}(\mathbf{r},\mathbf{r}^\prime=\mathbf{r};\varepsilon) \}
\end{equation}
\begin{equation}
\label{ni}
n_i(\varepsilon;\mu)=-\frac{1}{\pi} \int_{v_i} d \mathbf{r} \sum_L 
Im \{G_{ii,LL}(\mathbf{r},\mathbf{r}^\prime=\mathbf{r};\varepsilon) \}
\end{equation}.

As it is shown in Refs.~\onlinecite{Janak}, \onlinecite{DFTKKRCPA1},
\onlinecite{DFTKKRCPA2}, 
the Hohenberg-Kohn density functional, Eq.~(\ref{gpotNmu}),
can be more conveniently rewritten within the MST formalism 
as the sum of a kinetic and a potential energy 
functionals, as follows:
 \begin{equation}
\label{varform}
\Omega(T=0,V,\mu)=T - \mu N + U
\end{equation}
where the above two contributions are given by the following
expressions: 
\begin{eqnarray}
\label{Ti}
T-\mu N =
-\int_{-\infty}^{\mu} d \varepsilon N(\varepsilon;\mu)
-\int_V d \mathbf{r} \rho(\mathbf{r};\mu) v^{eff}(\mathbf{r};\mu) 
\end{eqnarray}
\begin{eqnarray}
\label{Ualt}
U&=&\frac{e^2}{2}\int_V d \mathbf{r} \int_V d \mathbf{r}^\prime
\frac{\rho(\mathbf{r}) \rho(\mathbf{r}^\prime)}
{|\mathbf{r}-\mathbf{r}^\prime|} 
-\sum_{j} \int_V d \mathbf{r} \frac{e^2 Z_j \rho(\mathbf{r})}{|\mathbf{r}-\mathbf{R}_j|} 
\nonumber \\
&+&\int_V d \mathbf{r} \rho(\mathbf{r}) e^{XC}(\mathbf{r}, [\rho])
+\frac{e^2}{2}\sum_{ij \; (i \neq j)} \frac{Z_i Z_j}{R_{ij}}
\end{eqnarray}

The effective potential in Eq.~(\ref{Ti}), $v^{eff}(\mathbf{r})$, 
is specified by the Kohn-Sham equation,
\begin{eqnarray}
\label{effpot}
v^{eff}(\mathbf{r})&=&\int_{V} d \mathbf{r}^\prime 
\frac{e^2 \rho(\mathbf{r}^\prime)}{|\mathbf{r}-\mathbf{r}^\prime |}
-\sum_j \frac{e^2 Z_j}{|\mathbf{r}-\mathbf{R}_{j}|} \nonumber \\
&+&v^{XC}(\mathbf{r}, [\rho])
\end{eqnarray}
It consists of 
the Coulombian potential due to the electronic and ionic charges and of the 
exchange-correlation potential, 
$v^{XC}(\mathbf{r}, [\rho])=\delta E^{XC}[\rho] / \delta \rho(\mathbf{r})$,
where $E^{XC}[\rho]$ is the third term on the RHS of Eq.~(\ref{Ualt}).
In the local density approximation (LDA) $v^{XC}$ is
assumed to depend locally on the electronic density, i.e., 
$v^{XC}(\mathbf{r}, [\rho])=v^{XC}(\rho(\mathbf{r}))$ and 
$e^{XC}(\mathbf{r}, [\rho])=e^{XC}(\rho(\mathbf{r}))$. 

We wish to highlight a useful consequence of the above partitioning 
of the system volume. The density functional defined by 
Eq.~(\ref{varform}), which is, of course,
variational wrt the global charge density, $\rho(\mathbf{r})$, turns out 
to be variational also wrt the charge densities in each 
scattering volume $v_i$, in formulae,
\begin{equation}
\label{varfun}
\frac{\delta \Omega} {\delta \rho_i(\mathbf{r})}=0
\end{equation}
Furthermore, it is possible to show\cite{DFTKKRCPA2} that  
\begin{equation}
\label{pass4}
\frac{\delta U}{\delta \rho_i(\mathbf{r})}=
v_i^{eff}(\mathbf{r};\mu)
\end{equation}
and that
\begin{equation}
\label{pass3}
\frac{\delta (T -\mu N)}{\delta \rho_i(\mathbf{r})}=
-v_i^{eff}(\mathbf{r};\mu)
\end{equation}

It is interesting to observe that, the expression for the site 
resolved DOS, Eq.~(\ref{ni}), allows to recast the integrated DOS 
and the electronic grand potential 
as sums of site resolved contributions. These contributions, however,
involve the site-diagonal part of the system Green function or scattering 
matrix, $\underline{G}_{ii}$, or $\underline{\tau}_{ii}$, and then 
are non trivially coupled together through the boundary 
conditions. If this coupling was neglected,
as it is done, for instance, in the case of the Harris-Foulkes density
functional\cite{Harris}, $O(N)$ scaling could be obtained. Fortunately, 
similar numerical performances can be achieved 
with less dramatic approximations. A sensible alternative is to impose
{\it random} boundary conditions at the fragments 
surfaces~\cite{Krajewski&Parrinello}. In this paper we shall follow a 
different approach and use {\it averaged} boundary conditions.
As we shall see in the following 
Section, this allows to have a tractable form for the coupling 
in the kinetic part of the density functional
and permit to obtain $O(N)$ algorithms. Although it was proposed with a
different aim, one of the oldest method applying such mean boundary
conditions is the CPA, a generalized version of which shall be offered in
the next Section. Before, however, we need to discuss a different 
source of coupling that is present in the potential energy part of the 
functional, namely, the electrostatic interactions between fragments.
In the past this subject has received little consideration and it has 
been ruled out by invoking the screening properties of
metals. However, nowadays there is a general consensus that careful 
estimates of these interactions are necessary in order to obtain accurate
total energies for metallic alloys. 

\subsection{Multipole expansions for the effective potentials and the potential energy}
\label{SectIIB}
We have shown in the previous Sect.~\ref{SectIIA} that the DF-MST theory is 
variational wrt the 
local charge densities, $\rho_i(\mathbf{r})$, of each fragment or scattering
volume. In this subsection we shall see how the multipole expansion  
used by most numerical implementations of the theory has the conceptual advantage
of giving expressions for the effective Kohn-Sham potentials and the potential 
energy in which different scattering volumes are coupled together only through
simple functions of the multipole moments. 

The relevant formulae can be obtained by splitting the volume integrals in Eqs.~(\ref{Ualt}) 
and~(\ref{effpot}) in sums of integrals extending over the scattering volumes, $v_i$ and  
by expanding the denominators in
spherical harmonics. Although they require some labour, the derivations are very 
straightforward and need not to be reported here. The resulting expressions for the potential 
energy and the effective potentials are listed below:
\begin{equation}
\label{Ualtsm}
U=\sum_i \left[ u_i(\rho_i(\mathbf{r}))+\frac{e^2}{2} \; \sum_L q_{i,L} V^{MAD}_{i,L} \right]
\end{equation}
and
\begin{eqnarray}
\label{effpotsm}
v_i^{eff}(\mathbf{r})&=&e^2 \int_{v_i} d \mathbf{r}^\prime 
\frac{\rho_i(\mathbf{r}^\prime)}{|\mathbf{r}-\mathbf{r}^\prime |}
-\frac{e^2 Z_i}{r}+ v^{XC}(\rho_i(\mathbf{r})) \nonumber \\
&+& e^2 V^{MAD}_i(\mathbf{r}) 
\end{eqnarray}

In Eqs.~(\ref{Ualtsm}) and~(\ref{effpotsm}) we have introduced the local 
multipole moments,
\begin{equation}
\label{qim}
q_{i,L}=\int_{v_i} d \mathbf{r} p_L(\mathbf{r}) \rho_i(\mathbf{r}) - Z_i \delta_{L,(0,0)} 
\end{equation}
and the Madelung potentials,
\begin{equation}
\label{vmadm}
V^{MAD}_i(\mathbf{r}) = \sum_L V^{MAD}_{i,L} p_L(\mathbf{r})
\end{equation}
where
\begin{equation}
\label{vmadm2}
V^{MAD}_{i,L}=\sum_{j \ne i} \sum_{L^\prime} M_{ij,LL^\prime} 
q_{J,L^\prime}
\end{equation}
The coefficients $M_{ij,LL^\prime}=M_{LL^\prime}(\mathbf{R}_{ji})$
are given by
\begin{equation}
\label{madgen}
M_{LL^\prime}(\mathbf{R})=4 \pi \sum_{L^{\prime\prime}\; (\ell^{\prime\prime}=\ell+\ell^\prime)} 
C_{LL^\prime}^{L^{\prime\prime}} \frac{(2\ell^{\prime\prime}+1)!!}{(2\ell^{\prime\prime}+1)}
\frac{Y_{L^{\prime\prime}}(\hat{\mathbf{R}})}{R^{\ell^{\prime\prime}+1}} 
\end{equation}
$C_{LL^\prime}^{L^{\prime\prime}}$ are the Gaunt numbers~\cite{Condon&Shortley}
and the functions $p_L(\mathbf{r})$ in Eq.~(\ref{effpotsm}) are defined as,  
\begin{equation}
\label{pdef}
p_L(\mathbf{r})=\frac{\sqrt{4 \pi}}{(2 \ell +1)!!} r^\ell Y^*_L(\mathbf{r})
\end{equation}
The only values that are relevant for spherical approximations are 
$p_{00}(\mathbf{r})=1$ and $M_{00,00}(R)=1/R$.

In Eq.~(\ref{Ualtsm}), the contribution from the $i$-th lattice site 
to the potential energy 
is denoted as $u_i([\rho_i(\mathbf{r})])$ and given by
\begin{widetext}
\begin{equation}
\label{Ualts2}
u_i([\rho_i(\mathbf{r})]) = 
\frac{e^2}{2}\int_{v_i} d \mathbf{r} \int_{v_i} d \mathbf{r}^\prime
\frac{\rho_i(\mathbf{r}) \rho_i(\mathbf{r}^\prime)}
{|\mathbf{r}-\mathbf{r}^\prime|}
 - \int_{v_i} d \mathbf{r} \frac{e^2 Z_i \rho_i(\mathbf{r})}{r}  
+\int_{v_i} d \mathbf{r} \rho_i(\mathbf{r}) e^{XC}(\rho_i(\mathbf{r}))
\end{equation}
\end{widetext}
Within the LDA, $u_i$ depends on the electronic density at the $i$-th
fragment only, while for non-local approximations
to the DFT there could be some dependence on the density at sites $j \ne i$. 

Much published work has been done within spherical 
approximations (SA), namely the muffin-tin (MT) or the atomic sphere approximation (ASA).
In that case only the first terms, $\ell=0$, of the multipole expansions are included. 
Thus, Eqs.~(\ref{Ualtsm}) and~(\ref{effpotsm}) must be replaced by
the following expressions,
\begin{equation}
\label{Ualts}
U=\sum_i \left[ u_i([\rho_i(\mathbf{r})])+\frac{e^2}{2} \; q_i V^{MAD}_i \right]
\end{equation}
and 
\begin{equation}
\label{effpots}
v_i^{eff}(\mathbf{r})=\int_{v_i} d \mathbf{r}^\prime 
\frac{e^2 \rho_i(\mathbf{r}^\prime)}{|\mathbf{r}-\mathbf{r}^\prime |}
-\frac{e^2 Z_i}{r}+v^{XC}(\rho_i(\mathbf{r}))+e^2 V^{MAD}_i
\end{equation}

Thus, for SA's, the only relevant multipole moments are the local charge excesses,
\begin{equation}
\label{qi}
q_i\equiv q_{i,00}=\int_{v_i} d \mathbf{r} \rho_i(\mathbf{r}) - Z_i 
\end{equation}
and the Madelung potentials are constant within each scattering volume to the values
\begin{equation}
\label{vmad}
V^{MAD}_i \equiv V^{MAD}_{i,00}=\sum_{j \ne i} \frac{q_j}{R_{ij}}
\end{equation}

Remarkably, in Eqs.~(\ref{Ualtsm}) and~(\ref{effpotsm}), or in their SA counterparts,
Eqs.~(\ref{Ualts}) and~(\ref{effpots}), the charge densities
at different sites, $\rho_i(\mathbf{r})$, are coupled only with the Madelung potentials 
at the same sites, $V^{MAD}_i(\mathbf{r})$. Of course, the last quantities contain information 
about the charge densities at all crystal sites. 

\begin{figure}
\includegraphics[width=7cm]{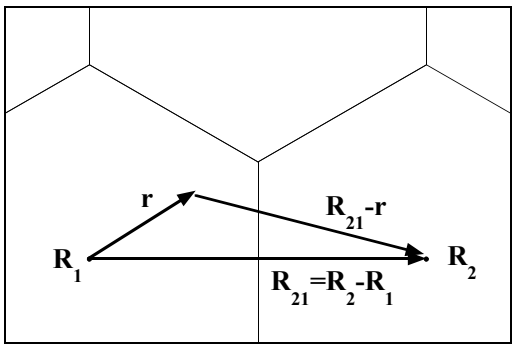}
\caption{Triangular inequalities that must be satisfied in order to have a
convergent multipole
expansion: $r<|\mathbf{r}-(\mathbf{R}_2-\mathbf{R}_1)|$, $r<|\mathbf{R}_2-\mathbf{R}_1|$.
The partition of the system volume in Voronoi polyhedra, marked by the lines, 
guarantees that the inequalities hold.}
\label{triang}
\end{figure}

We wish to highlight that the multipole expansion does not converge
for arbitrary partitions of the system. Actually, convergence requires
that, for any pair of scattering centers, $\mathbf{R}_i$ and $\mathbf{R}_j$, 
and for any point $\mathbf{r}$ belonging to the scattering volume $v_i$,  
the triangular inequality illustrated in
Fig.~(1) must be satisfied.
It is easy to realize that partitioning of the system in Voronoi polyhedra accordingly
with the Wigner-Seitz construction guarantees the above condition to be fullfilled
everywhere except but for the zero  measure set of points constituted by the surfaces
of the polyhedra, thus ensuring the convergence of the theory.
The Wigner-Seitz construction, therefore, constitutes a natural choice
for the partitioning.

\section{Generalized Coherent Potential Approximations (GCPA) and
Charge Excesses Functional theory (CEF)}
\subsection{Generalized Coherent Potential Approximations (GCPA) for the 
scattering matrices}
\label{GCPA}
In this Section we shall discuss a whole class of approximations for systems
with atoms lying on a regular lattice, where, however, substitutional
disorder is allowed for. Metallic alloys, both ordered intermetallic 
compounds and random alloys, constitute the most relevant example of such systems. 
Other examples are crystals with empty, or 'vacancy', sites. Although, 
in general, these systems do not have translational invariance,
nevertheless the underlying 'geometrical lattice' does. Forty years ago, 
this consideration led Soven to formulate the Coherent Potential Approximation 
or CPA~\cite{soven}. Since then, the CPA had an appreciable fortune. Its 
crucial virtue was that, by introducing a 'mean field' fashion 
effective crystal, it allows to use many techniques designed for ordered 
systems that were already well developed  at the time at which the theory was
proposed.

For many years, the DFT implementations of the CPA~\cite{F&S,DFTKKRCPA1} 
have been based on the assumption (in the following referred to as the 
single-site approximation or SS) that sites occupied by atoms of the same chemical 
species are characterized by the same effective Kohn-Sham potentials. Although the 
DFT-SS-CPA has been proved able to carefully determine the electronic 
structure
and the spectral properties of many alloy systems~\cite{Faulknerrev,
Abrikosov_cpa,CPALF}, nevertheless it leads to an incorrect description 
of the electrostatics and of
the total energies in metallic alloys~\cite{Magri}. Due to its mean field nature,  
in fact, the SS approximation neglects the fluctuations of the charge 
transfers and the energetic electrostatic 
contributions associated with them. This failure
has stimulated many authors that envisaged CPA generalizations  
aimed to include the effects of different chemical 
environments~\cite{isomorphous,Ujfalussy,SIMCPAI,SIMCPAII,ccCPA}. 

In this paper we define a {\it class of approximations} for DFT-based 
electronic theories in which most of the above CPA generalizations can be 
included. We shall refer to the approximations belonging to
such a class to as Generalized CPA (GCPA). A theory belonging to the GCPA class
shall be identified by: (a) a theory specific {\it 'external model'}, i.e. 
a rule for determining the effective Kohn-Sham 'site' potentials and 
the statistical weights $w_i$ to be assigned to each 'site', and
(b) an approximate form for the kinetic part of the density functional, 
specified by Eqs.~(\ref{CPA0}-\ref{CPA3}) below. The last feature is common 
to all the theories belonging to the GCPA class.
  
Before discussing the ansatz for the kinetic functional we wish to illustrate
what a GCPA 'external model' can be on the basis of a few examples.
The first example of a GCPA theory is, of course, the DFT implementation of the 
SS-CPA in Refs.~\onlinecite{DFTKKRCPA1} and ~\onlinecite{DFTKKRCPA2}. Its 
external model is the SS 
assumption (identical effective potentials for atoms of the same atomic species and 
weights proportional to the respective atomic concentrations). 
Another example is the Polymorphous CPA (PCPA) of Ujfalussy et al.~\cite{Ujfalussy,
Faulkerphilmag,PCPA2001}. 
The external model is constructed using an auxiliary supercell 
containing $N$ atoms, usually hundreds or thousands, each to be 
weighted with the same weight. The effective site
potentials are reconstructed on the same
supercell via Eq.~(\ref{effpot}), thus atoms of the same chemical species 
are allowed to have different potentials depending on their environments. 
This specific choice for the external model appears the reason why the
PCPA theory substantially improves the alloy 
electrostatics while maintaining all the advantages of the standard SS-CPA 
about the spectral properties~\cite{PCPAapplications}. Other existing
CPA-based approaches like, e.g., the Non-Local CPA~\cite{NLCPAI,NLCPAII},
or the SIM-CPA~\cite{SIMCPAI,SIMCPAII} can also be considered as particular
cases of GCPA's.
  
We shall now introduce the kinetic ansatz that is common to all GCPA
theories. For this purpose we prefer not to start from the definition of
the functional. Rather we shall follow a path closer to physical 
intuition and to the spirit
of Soven's original CPA formulation\cite{soven}.
At similarity of SS-CPA calculations, the GCPA defines an effective 
periodic crystal whose sites are occupied by effective 'coherent' scatterers 
characterized by the single-site scattering matrix 
$\underline{t}^c(\varepsilon)$,
the corresponding Green function shall be 
$\underline{G}^c(\underline{t}^c)$. 
Then, if we considers the Green function of a single 
substitutional impurity with a single-site scattering matrix $\underline{t}_i$ 
embedded in the above effective crystal, 
$\underline{G}_{ii}(\underline{t}_i,\underline{t}^c)$, the GCPA consists 
in requiring that
\begin{equation}
\label{CPA0}
\sum_i \frac{w_i}{N} \; \underline{G}_{ii}(\underline{t}_i,\underline{t}^c) =
\underline{G}^c(\underline{t}^c)
\end{equation}
In other words, the weighted average of the impurity Green functions
must be equal to the 'coherent' Green function
$\underline{G}^c(\underline{t}^c)$. In Eq.~(\ref{CPA0})  
the energy dependences have been dropped for sake of simplicity and $N$
stands for the number of different scatterers in the model.

\begin{figure}
\includegraphics[width=7cm]{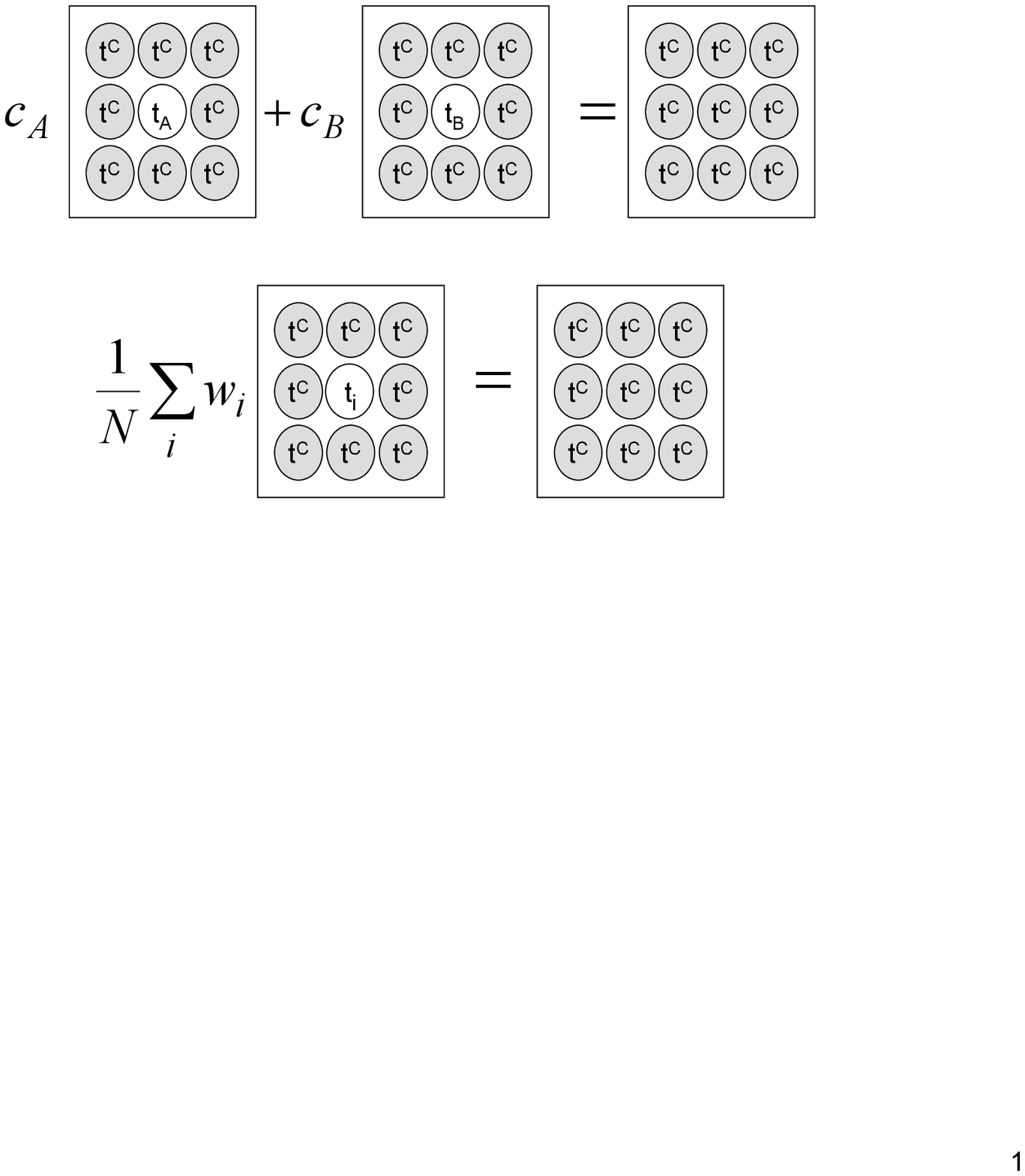}
\caption{\label{fig2} A pictorial illustration of the SS-CPA for
a binary, A$_{c_A}$B$_{c_B}$, alloy (top) and of the GCPA
(bottom). The rectangular frames have the meaning "the Green function 
of what is inside".}
\end{figure}

Eq.~(\ref{CPA0}) is illustrated in Fig.~(2).
In terms of the 'coherent'
scattering-path matrix of the effective lattice, $\underline{\tau}^c$,  
and of the CPA 'projectors', $\underline{D}_i$, it can be 
rearranged as follows:
\begin{equation}
\label{CPA1}
\sum_i \frac{w_i}{N} \underline{D}_i =\underline{1}
\end{equation}
\begin{equation}
\label{CPA2}
\underline{D}_i = \left[\underline{1}+\left(\left(\underline{t}_i\right)^{-1}-
\left(\underline{t}^c\right)^{-1}\right) 
\underline{\tau}^c \right]^{-1}
\end{equation}

A GCPA theory is then an approximation for the $\underline{\underline{\tau}}$ 
matrix, whose diagonal elements are given by
\begin{equation}
\label{CPA3}
\underline{\tau}_{ii}=\underline{D}_i \; \underline{\tau}^c
\end{equation}
while the diagonal matrix elements of the Green function are given by
Eq.~(\ref{greenfun}) with $i=j$.
  
Within the approximation defined by Eqs.~(\ref{CPA0}-\ref{CPA3}) above, 
the MST reviewed in the previous Section allows to calculate the charge
densities and the integrated DOS, $N(\varepsilon,\mu)$ and, through
Eq.~(\ref{Ti}), the kinetic part of the Hohenberg-Kohn functional. Here
we need not to trace all the intermediate steps that can be
reproduced following the scheme of Ref.~\cite{DFTKKRCPA2}. The GCPA
approximate version of the Lloyd formula, Eq.~(\ref{MKKR}), is given by:
\begin{eqnarray}
\label{Lloyd3}
\frac{N(\varepsilon;\mu)}{N} 
&=&\frac{N^0(\varepsilon)}{N}+\frac{1}{\pi} 
Im \sum_i w_i Tr \ln \underline{\tau}^c(\varepsilon) \nonumber \\
&+&\frac{1}{\pi} Im \sum_i w_i Tr \ln \underline{D}_i(\varepsilon)
\end{eqnarray}
It has the very remarkable property  
that the integrated DOS $N(\varepsilon;\mu)$ and, hence, the kinetic functional 
are variational~\cite{DFTKKRCPA2} wrt. both $\underline{t}^c$ and 
$\underline{\tau}^c$. 

In Sec. II A, we have mentioned that in the exact MST 
the contributions to the integrated DOS associated with 
each lattice site
and proportional to $(ln \underline{\tau})_{ii}$, are coupled together
because each element of the $\underline{\underline{\tau}}$ matrix depends on 
the scattering properties of all the lattice sites. Within the GCPA, the 
only source of coupling is $\underline{\tau}^c$. Each local contribution 
depends on $\underline{\tau}^c$ and on the local potential. However, within 
the GCPA,
the Lloyd formula does not depend on $\underline{\tau}^c$ nor on the 
local potentials.
As a consequence, the integrated DOS results in a sum of local 
contributions, coupled together only through $\underline{\tau}^c$,    
\begin{equation}
\label{Lloyd4}
N(\varepsilon;\mu) = \sum_i w_i N_i(\varepsilon;\mu)
\end{equation}
We shall call this very controlled and tractable kind of coupling 
{\it marginal coupling}.

In view of further developments, it is convenient to isolate in Eq.~(\ref{Lloyd4}) 
two distinct terms. 
The first arises from the first two addends in 
Eq.~(\ref{Lloyd3}), it is identical for all sites and related to the 
effective background 
defined by the GCPA medium. The second depends on the local CPA projectors 
and, through them, on the local potentials. In formulae:      
\begin{equation}
\label{Lloydterm}
N_i(\varepsilon;\mu) 
=N_i^{back}(\varepsilon;\mu)+\frac{1}{\pi} Im Tr \log \underline{D}_i(\varepsilon)
\end{equation}

Implementing the GCPA within the DFT gives for the kinetic functional of 
Eq.~(\ref{Ti}) the following marginally coupled form:
\begin{eqnarray}
\label{kin1st}
T-\mu N = T^{back}(\mu) +  \sum_i w_i T_i ([\rho_i],\mu)
\end{eqnarray}
where
\begin{equation}
\label{Tback}
T^{back}(\mu) = - \int_{-\infty}^\mu d\varepsilon N^{back}(\varepsilon;\mu)
\end{equation}
and
\begin{eqnarray}
\label{Tigcpa}
T_i([\rho_i],\mu)&=& \frac{1}{\pi} Im Tr \int_{-\infty}^\mu 
d\varepsilon \log \underline{D}_i(\varepsilon) \nonumber \\
&-&\int_{v_i} d \mathbf{r} \rho_i(\mathbf{r};\mu) v_i^{eff}(\mathbf{r};\mu) 
\end{eqnarray}

As mentioned in Sect. II, in MST-based DFT calculations, 
the only source for $O(N^3)$ scaling is the inversion of the multiple 
scattering matrix, Eq.~(\ref{MKKR}), required
to obtain the scattering-path matrix $\underline{\underline{\tau}}$. 
This step is bypassed in a GCPA
theory by approximating the relevant matrix elements $\underline{\tau}_{ii}$ 
via Eq.~(\ref{CPA1}) in terms of the local scattering properties and
the coherent scattering matrix $\underline{\tau}^c$, the last of which 
is, in turn, obtained by 
an averaging process. For this reason, GCPA theories are $O(N)$,
allowing for very substantial savings of computing time. Of course,
the price for these savings is payed by the approximation implied by
Eq.~(\ref{CPA3}). A diagrammatic analysis of these errors can be found
in Ref.~\onlinecite{Gonis}.

It is necessary to make a couple of remarks about the 
physical meaning of the GCPA in the present context and to highlight the 
differences with respect to the traditional way in which CPA-based 
theories have
been introduced in the past. In first place, the GCPA has been 
introduced here as
an approximation for the Hohenberg-Kohn density functional. As an
approximation, it may well be used to describe an ordered alloy. Its
range of applicability is by no means confined to the realm of random
alloys. In second place, the introduction of the weights $w_i$ to be
assigned to each scatterer makes GCPA theories suitable for dealing
with sophisticated pictures of the order (or the disorder) in metallic
alloys. This, of course, requires what we have called an 'external
model'. In a foregoing paper we shall discuss a (to some extent)
self-consistent way to define an external model that is able to provide
a picture of ordering phenomena in metallic alloys as a function
of the temperature.   

\subsection{The DFT-MST-GCPA functional: the 'marginal coupling' property}
In the present subsection we shall analyze certain formal properties of 
the GCPA approximations introduced in the previous subsection. 
All the above discussion can be summarized in the following approximate 
density functional:
\begin{eqnarray}
\label{gpot}
\Omega^{GCPA}&=&T^{back}(\mu)+\sum_i w_i 
\bigg[ \omega^{GCPA}_i([\rho_i],\mu) \nonumber \\
&+&\frac{e^2}{2} \sum_L q_{i,L}  V^{MAD}_{i,L} \bigg] 
\end{eqnarray}
In Eq.~(\ref{gpot}) the $q_{i,L}$ are defined by Eq.~(\ref{qim})
and the {\it local part} of the GCPA {\it functional} by
\begin{equation}
\label{gpoti}
\omega^{GCPA}_i([\rho_i],\mu)= T_i([\rho_i],\mu)+u_i([\rho_i],\mu)
\end{equation}
where the terms $T_i([\rho_i],\mu)$ and $u_i([\rho_i],\mu)$ are given by 
Eqs.~(\ref{Tigcpa}) and~(\ref{Ualts2}) above. We note that the local GCPA 
functional, $\omega^{GCPA}_i$, depends also on the atomic number 
of the atom at $\mathbf{R}_i$, $Z_i$, and on the volume and the shape of 
the $i-th$ 
Voronoi polyhedron through the local potential energy term, $u_i$. 
In the following we shall make the simplifying assumpion
of having identical Voronoi polyhedra for all the sites considered.

In Eq.~(\ref{gpot}) the coupling potentials, $V^{MAD}_{i,L}$, are provided by 
the specific external model. In the following of this Section we shall 
assume:
\begin{equation}
\label{vmadlam}
V^{MAD}_{i,L}= \sum_{j \ne i} \sum_{L^\prime} \lambda_L \lambda_{L^\prime} 
w_j M_{ij,LL^\prime,} q_{j,L^\prime}
\end{equation}
Appropriate choices of the coefficients $\lambda_L$ and of the weights, 
$w_i$, give then the SS-CPA or the PCPA. Furthermore, Eq.~(\ref{vmadlam}) 
can also be used for spherical or for full-potential charge reconstructions.   

As mentioned in Sect. II A, Eq.~(\ref{varfun}), the density functional 
is variational not only wrt the global charge density, $\rho(\mathbf{r})$, 
and the chemical potential $\mu$, 
but also wrt the charge densities in each scattering volume, 
$\rho_i(\mathbf{r})$. Moreover, as discussed in Sec. III A
and in Ref.~\onlinecite{DFTKKRCPA2}, 
the GCPA density functional is variational wrt the effective medium 
scattering matrix, $\underline{\tau}^c$. Furthermore, in a GCPA theory, 
the background 
kinetic term, $T^{back}(\mu)$ in Eq.~(\ref{gpot}) depends on the electronic
density only through $\underline{\tau}^c$ and $\mu$. Thus, the functional 
derivation of
Eq.~(\ref{gpot}) wrt the local densities, $\rho_i(\mathbf{r})$, gives the
following set of coupled equations:
\begin{equation}
\label{varGCPA}
\frac{\delta \omega^{GCPA}_i}{\delta \rho_i(\mathbf{r})} +
V^{MAD}_i(\mathbf{r}) =0 
\end{equation}
where we have used Eqs.~(\ref{qim}), (\ref{vmadm}), (\ref{vmadm2}) and~(\ref{pdef}). 

Within a GCPA theory, solving the set of the Euler-Lagrange 
equations~(\ref{varGCPA}), one for each 
scattering center, together with the equations that determine the chemical 
potential and the coherent scattering matrix $\underline{\tau}^c$, is completely
equivalent to the minimization of the density functional. As it is
apparent, these Euler-Lagrange equations are coupled each other {\it only}
through the Madelung potentials, $\underline{\tau}^c$ and $\mu$. 
Moreover, the
functionals $\omega^{GCPA}_i(\rho_i(\mathbf{r}))$ are identical for sites 
occupied by the same chemical species.

In order to understand the consequences of the above result let us consider, 
for instance, an alloy sample constituted by a large supercell. 
One may wish to calculate, {\it in the given sample}, the 
properties of different 'atoms', in first place the charge densities, 
$\rho_i(\mathbf{r})$. Inside the sample, $\underline{\tau}^c$, $\mu$ 
and the cell geometry are fixed, thus the set of the $\rho_i(\mathbf{r})$
is completely determined by the values of the Madelung potentials
$V^{MAD}_i(\mathbf{r})$ and by the atomic number $Z_i$ of the ion at the 
position
${\mathbf{R}_i}$. More generally, inside the given sample, any site diagonal 
property $\Pi_i$ shall be completely determined by $Z_i$ and by the set of Madelung
potentials. We can establish this result as follows:
\begin{equation}
\label{Pii}
\Pi_i=\Pi(Z_i,V_i^{MAD}(\mathbf{r}))
\end{equation}  
Examples of such site diagonal properties are the local contributions to 
the grand potential, the multipole moments, and the local DOS.
The functional forms, one for each alloying species, given by 
Eqs.~(\ref{Pii}) sometimes can be easily numerically fitted and then 
constitute a useful tool for the evaluation
of the site quantities in the given sample. They are the source
of the simple laws, as e.g. the 'qV' laws, empirically found 
from extended metallic systems calculations. This notwithstanding,
GCPA theories are able to predict complex trends for certain
site diagonal properties as, e.g., the site resolved 
DOS's~\cite{PCPAapplications,PCPADOS}. 

Since Eqs.~(\ref{Pii}) allows to evaluate, among other properties, also 
the charge density of each
fragment and, hence, the full charge density, $\rho(\mathbf{r})$, then, in 
virtue of the Hohenberg and Kohn theorem, it follows that any ground 
state observable
in the sample given is a {\it functional} of the set of
the Madelung potentials at all the crystal sites, 
$V^{MAD}_i(\mathbf{r})$, and of the set of the atomic numbers {\it only}.
Since the last is, again, specified by the sample, it follows the theorem:
{\it any
ground state observable in the sample given is a functional of the
Madelung potentials only}, or, equivalently, {\it a function of the set 
of coefficients},
$\{V^{MAD}\}$, that completely determine the Madelung potentials.
\footnote{
In our notation $\{V^{MAD}_i\}$ stands for the set of all the Madelung
coefficients of the $i$-th site, $\{V^{MAD}_{i,L_1}, V^{MAD}_{i,L_2}, 
\cdots\}$ and $\{V^{MAD}\}$ for the set of all the Madelung 
coefficients at all crystal sites, i.e. $\{V^{MAD}\}=\{V^{MAD}_{i_1}\}
\cup \{V^{MAD}_{i_2}\} \cup \cdots$. The notations $\{q_i\}$ and $\{q\}$
have a similar meaning.}
 
Since, in virtue of Eq.~(\ref{vmadlam}), the coefficients in the set 
$\{V^{MAD}\}$ are  
linear functions of the set of the multipole moments, $\{q\}$, 
then the above theorem implies the corollary that any ground state property
of the sample is a function of the same moments. Within the GCPA and
for the specific sample given, it is then possible 
to reformulate the DFT in terms of 
the charge multipole moments. By neglecting a constant term with the 
physical meaning
of the grand potential contribution due to the mean GCPA 'atom',
Eq.~(\ref{gpot}) can be written as 
\begin{eqnarray}
\label{gpotcg}
&&\widetilde{\Omega}^{GCPA}(\{q\},\mu)=\sum_i w_i 
\widetilde{\omega}^{GCPA}_i(\{q_i\}, Z_i) \nonumber \\
&&+\frac{e^2}{2} \sum_{i,j,L,L^\prime} w_i w_j  
\lambda_L  \lambda_{L^\prime}  
M_{ij,LL^\prime} q_{i,L} q_{j,L^\prime} \nonumber \\
&&- \mu \sum_i w_i q_{i,00}
\end{eqnarray}

In deriving Eq.~(\ref{gpotcg}), we have used Eq.~(\ref{vmadlam}) and 
the fact that, since $\omega^{GCPA}_i$ is
completely determined by the local density, $\rho_i(\mathbf{r})$, it cannot depend
on the multipole moments at other sites. Moreover, 
we have isolated the contribution proportional to the chemical
potential, $\mu$. Having introduced an explicit dependence on $\mu$,
the last term in Eq.~(\ref{gpotcg}) can be thought as a way of enforcing
the global electroneutrality. This is unnecessary if we consider a
specified sample in which, of course, $\mu$ has a precise, fixed vale.
However, since the term proportional to $\mu$ has precisely the form
it must have, its introduction is equivalent to extend the validity of 
Eq.~(\ref{gpotcg}) to {\it all samples specified by the same mean 
atomic concentrations and by the same value for} $\underline{\tau}^c$.

To summarize, we have established the following results. Within the
GCPA class of approximations the Hohenberg and Kohn density functional
can be recast in the form of Eq.~(\ref{gpotcg}). It consists of
(a) local terms, $\widetilde{\omega}^{GCPA}_i$ for the $i$-th 
scattering site, consisting in functions of the charge multipole moments
that are identical for sites with the same chemical occupation; (b) a 
bilinear form coupling the charge multipole moments at different sites,
with coupling coefficients $M_{ij,LL^\prime}$ defined
by the crystal geometry; (c) a term proportional to the chemical potential
that ensures the global electroneutrality. The functional defined by
Eq.~(\ref{gpotcg}) is identical for all the alloy samples characterized 
by the same mean atomic concentrations and the same value for the coherent
scattering-path matrix $\underline{\tau}^c$. Evidently, it
constitutes a {\it coarse grained} version of the DTF because the 
mathematical definition of the multipole moments, Eq.~(\ref{qim}), 
does not completely determine the charge density. The last is
determined by the multipole moments only within the GCPA theory. 
This reduction of the relevant information has been obtained at the 
price (a) of the GCPA approximation and (b) of having restricted the 
consideration to a specific sample. Nevertheless, no restriction has 
been made about the size of the 
sample that, therefore can be chosen in such a way to guarantee an 
appropriate description for a fixed concentration ensemble, as we shall 
discuss at the end of the present section. 

Having recast the GCPA {\it functional} as a sum of {\it functions} of
the charge multipole moments has obvious mathematical advantages. However,
we have not yet completely determined the functional form of the local
energetic contributions $\widetilde{\omega}^{GCPA}_i(\{q_i\},Z_i)$. 
In order to do this, we need to make the hypothesys that, in the sample
considered, the distribution of the Madelung potentials coefficients,
$\{V^{MAD}\}$, is continuous in the range of the values that the same 
potentials assume in the sample. This is consistent with the observations
in Refs.~\onlinecite{FWS1}, \onlinecite{FWS2}, \onlinecite{CEF}. 
Let us consider two scattering sites, say $i$
and $j$, occupied by the same chemical species, $\alpha$, at which
the Madelung coefficients take very close numerical values, 
$V^{MAD}_{i,L}=V^{MAD}_L$ and $V^{MAD}_j(\mathbf{r})=V^{MAD}(\mathbf{r})+\Delta V^{MAD}(\mathbf{r})$.
The local energetic contributions, the charge densities and the local
multipole moments shall be: 
$\tilde{\omega}^{GCPA}_i=\tilde{\omega}^{GCPA}_\alpha$, 
$\rho_i(\mathbf{r})=\rho(\mathbf{r})$ and $q_{i,L}=q_L$ for the $i$-th
site, and 
$\tilde{\omega}^{GCPA}_j=\tilde{\omega}^{GCPA}_\alpha+\Delta \tilde{\omega}^{GCPA}_\alpha$, 
$\rho_j(\mathbf{r})=\rho(\mathbf{r})+\Delta \rho(\mathbf{r})$ and 
$q_{j,L}=q_L+\Delta q_L$ for the $j$-th site. To the first order in 
$\Delta \rho(\mathbf{r})$ we have:
\begin{eqnarray}
\Delta \tilde{\omega}^{GCPA}_\alpha & = &\int_{v_i=v_j} d \mathbf{r}
\bigg( \frac{\delta \tilde{\omega}^{GCPA}_\alpha}{\delta \rho(\mathbf{r})}
\bigg)_{\rho_i(\mathbf{r})=\rho(\mathbf{r})} \Delta \rho(\mathbf{r}) 
\nonumber \\
& = &- \int_{v_i=v_j} d \mathbf{r} V^{MAD}(\mathbf{r})
\Delta \rho_(\mathbf{r}) \nonumber
\end{eqnarray}
where Eq.~(\ref{varGCPA}) has been used. The substitutions of the expansion
for the Madelung potential, Eq.(\ref{vmadm}), and of the expressions for 
the charge multipole moments, Eq.(\ref{qim}), then give:
\begin{equation}
\label{domega}
\Delta \tilde{\omega}^{GCPA}_\alpha = -\sum_L V^{MAD}_L \Delta q_L
\end{equation}
Once integrated over $q_L$ Eq.~(\ref{domega}) gives
\begin{equation}
\label{intomega}
\tilde{\omega}^{GCPA}_\alpha(\{q\}) =\tilde{\omega}^{GCPA}_\alpha(\{q^0\})
-\sum_L \int_{\{q^0\}}^{\{q\}} 
V^{MAD}_{L,\alpha}(\{q\prime\}) d q_L^\prime
\end{equation}
Eq.~(\ref{intomega}) can be easily numerically evaluated from the 'qV' data,
$V^{MAD}_{L,\alpha}=V^{MAD}_{L,\alpha}(\{q\})$ 
obtained as an output from GCPA calculations. Unless a constant with the 
meaning of the local energy contribution at $\{q\}=\{q^0\}$, it determines
the local energies for each chemical species $\alpha$. 
Eqs.~(\ref{domega}) and~(\ref{intomega}) have 
been obtained under very broad conditions: the differentiability of the 
kinetic functional~\cite{lieb,Englisch} and the monotonicity of the 'qV' 
laws. The first is
the usual requirement for the convergence of the Kohn-Sham scheme of the DFT,
while the second condition is certainly verified by all GCPA calculations
reported in the literature, including those executed at extremely high 
values for the Madelung potential (see Fig. 7 in the next subsection and 
the related discussion). 

In the remainder of the present Section we shall make a few general comments
about the validity of the framework defined by the GCPA theory in comparison
with the exact density functional.

(i) The fact that the effective potential and the potential energy functional 
can be decomposed in site contributions coupled together only through 
the Madelung potentials is an exact consequence of the LDA 
and it has nothing to do with the GCPA. This kind of coupling is 
{\it 'marginal'} in the sense that, although not necessarily small,
it has the simple and tractable functional form which arises 
from the bilinear terms involving the multipole moments in Eq.~(\ref{gpotcg}).
Although this is beyond the purpose of the present paper, we notice that
most non-local density functionals offered in the 
literature~\cite{Dreizler&Gross,Dreizler&daProvidencia} 
are actually local wrt the density gradients, thus most non-local schemes
will remain {\it marginally} coupled in the above sense.

\begin{figure}
\includegraphics[width=6.8cm,height=3.06cm]{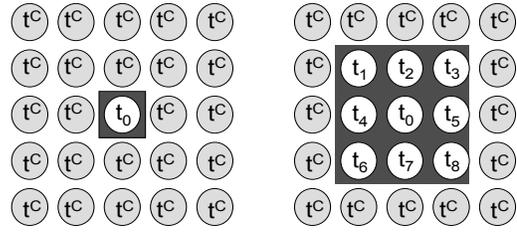}
\caption{\label{comparison} The Local Interaction Zones (LIZ) used in PCPA
(left frame) and LSGF (right frame) calculations are marked by the dark 
areas. For each scattering site, the kinetic functional is evaluated 
using the appropriate single-site scattering matrices, $t_i$, 
inside the LIZ, 
and the effective CPA scattering matrix, $t^c$, outside.}
\end{figure}

(ii) Splitting the kinetic functional into local contributions
marginally coupled trough the coherent scattering matrix 
$\underline{\tau}^c$ is a simplification
due to the GCPA. In fact, it has been obtained by
assuming averaged boundary conditions at the surfaces of
the Voronoi polyhedra through Eq.~(\ref{CPA0}). An
estimate of the so induced errors can be obtained by the comparison of
PCPA vs. Locally Self-Consistent Green
Function (LSGF) calculations~\cite{LSGF} executed on the same supercell. 
As it is sketched in Fig.~(3), both calculations
evaluate the kinetic contribution from the $i$-th to the functional by   
solving the problem of a single impurity, in the case of the PCPA, or
of an impurity cluster, the Local Interaction Zone (LIZ), for the LSGF. 
In both cases the scattering matrices outside the LIZ are set to the
coherent scattering matrix, $\underline{t}^c$. PCPA calculations can 
then be viewed as LSGF calculations with only one atom in the LIZ. This
argument also suggest that, wrt GCPA calculations, exact DFT results
include, for each site, corrections depending on its chemical environment.  

(iii) We have already seen that the coarse grained version of the GCPA 
functional, Eq.~(\ref{gpotcg}), 
holds for all the alloy configurations characterized by a specified value 
for $\underline{\tau}^c$ in a fixed concentration ensemble.
This could appear as a serious limitation since it looks   
unlikely that, e.g., $\underline{\tau}^c$ could have the same functional 
energy dependence for two different systems.  
In general, in a GCPA theory, $\underline{\tau}^c$ is a ground
state property determined not only by the mean concentrations 
but also by the distributions of the Madelung potentials for each alloying 
species. As opposite to the SS model where these distributions are trivial, 
more sophisticated external models, as e.g. the PCPA, give complicated
charge and Madelung potential distributions. How could then the GCPA 
functional be useful in such cases?
\begin{figure}
\includegraphics[width=7cm]{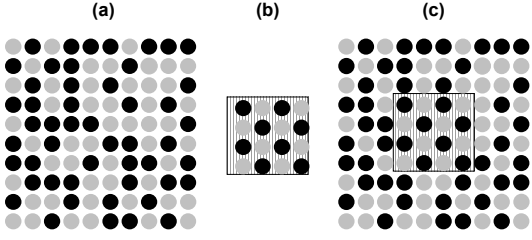}
\caption{The PCPA supercells shown in frame (a) and (b) contains, 
respectively, $N \rightarrow \infty$ atoms in a random alloy configuration
and $n$ atoms in an ordered configuration, both at the same mean atomic
concentrations. $N$ is large enough to to guarantee an appropriate description 
of a random alloy within a GCPA theory. Similarly, $n$ has been chosen to 
permit the description of an ordered alloy up to some length scale $l$. 
The supercell in the frame (c) is identical to that in (a) except but for the 
dashed region which contains $n$ atoms in the same ordered configuration
as in (b). The cell (c) is therefore able to describe an ordering fluctuation
up to the scale set by $l$.
}
\end{figure}
As argued by Faulkner et al.~\cite{PCPAmath}, the PCPA theory applied to 
ideal random alloys gives well-defined values for all physical properties. 
This is because, in perfect random alloys, the distribution of the 
chemical environments is easily obtained by statistical 
considerations and it is given by the appropriate multinomial distributions.
Therefore, the PCPA random alloy constitutes a privileged reference
system whose physical properties, including $\underline{\tau}^c$,
can be approximated up to an arbitrary accuracy by letting the 
number of atoms in the PCPA supercell, $N$, going to infinity (see Fig. 4a). 
We believe that the same $\underline{\tau}^c$ obtained for a random alloy 
at a given concentration can be used for building a physically clean, 
though approximate, theory also for ordered alloys at the same 
concentration. In the next section we shall provide numerical evidences
for that, here we present a more formal argument. Imagine that
an ordered array containing $n$ atoms (Fig.4b) is able to account for 
the properties of same ordered alloy configuration, up to a length scale, 
$l$, that can be made large at will in the $n \rightarrow \infty$ limit. 
In Fig. 4c we draw a supercell, a part of which is constituted by the
supercell of Fig. 4b, while the remaining $N-n$ sites are occupied as
in the random alloy supercell of Fig. 4a. We can think that the supercell 
in Fig. 4c represents a fluctuation of an ordered phase in a random alloy 
matrix and that it describes the physical properties 
of such fluctuation up to the same length scale $l$ as in Fig. 4b. 
We are implicitly using the common idea of 'locality' in physics,
or, in a more specific context, of 'nearsightedness' of the 
DFT\cite{nearsightedness}. However, 
as Eq.~(\ref{CPA1}) implies, the difference between the coherent 
scattering-path matrices corresponding to Figs. 4a and 4c, 
$\underline{\tau}^c_a - \underline{\tau}^c_c$,
is proportional to the ratio $n/N$ and, then, it can be made small at will
in the $N \rightarrow \infty$ limit, for {\it any} value of $n$. We conclude
that the coherent scattering-path matrices of a random alloy,
$\underline{\tau}^c_a$, is able to 
account for the physical properties of ordered configuration considered.
The limitation
$n/N <<1$, that comes from the above argument, does not impose any
upper bound on the maximum length scale at which chemical fluctuations 
can be studied and it is of no practical importance provided that $N$
is large enough to ensure a good approximation $\underline{\tau}^c_a$.
As reported in the literature~\cite{PCPAapplications,Faulkerphilmag}, 
it seems that $N$ about 100 is already enough.

iv) Although we have suggested that the coherent scattering matrix from 
random alloys GCPA calculations can be used for ordered alloys too, we are
aware of the limitations of such a physical picture. For instance, a GCPA 
theory always implies finite quasiparticle lifetimes~\cite{physrep},
and, hence, a smearing of the peaks of the Bloch spectral function (BSF).

\subsection{A generalized version of the Charge Excess Functional Theory}

As a matter of fact, the analysis of DFT supercell calculations 
for metallic alloys suggests the existence of simple relationships
between the charge excesses at the lattice sites, $q_{i,00}$, and
the Madelung potentials at the same sites, $V^{MAD}_{i,00}$. Namely,
simple linear laws, one for each alloying species, have been found 
to hold, say
\begin{equation}
\label{qv}
a_i q_{i,00} + V^{MAD}_{i,00} = k_i
\end{equation}
where $a_i$ and $k_i$ have the same numerical values for atoms
of the same chemical species in the given supercell. 
Examples of the linear 'qV' laws obtained from PCPA calculations
for a binary and a quaternary alloy are reported in Figs. 5 and 6.

\begin{figure}\label{binqV}
\includegraphics[width=7cm]{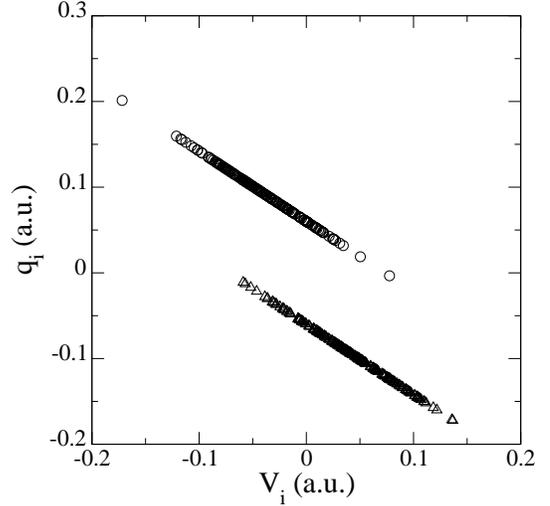}
\caption{'qV' relationships for a bcc random $Cu_{0.50}Zn_{0.50}$ alloy.
The excesses of electrons, $q_i=q_{i,00}$ are plotted vs. the Madelung potentials, 
$V^{MAD}_i$. The results have been obtained from $\ell_{MAX}=0$ PCPA 
calculations for a supercell containing 432 atoms at lattice costant $a=5.50$ a.u.. 
Circles represent Cu atoms and triangles Zn atoms. Note that positive values for
$q_i$ correspond to negative net charges and vice versa.
}
\end{figure}

\begin{figure}
\includegraphics[width=7cm]{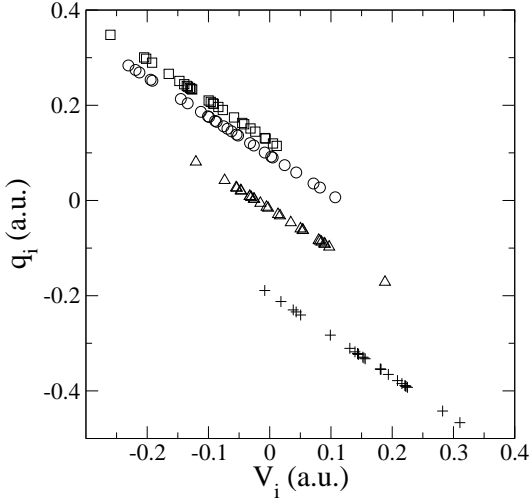}
\caption{\label{quaternqV}
'qV' relationships for an fcc random $Al_{0.25}Cu_{0.25}Ni_{0.25}Zn_{0.25}$ 
alloy. The excesses of electrons, $q_i=q_{i,00}$ are plotted vs. the Madelung potentials, 
$V^{MAD}_i$. The results have been obtained from $\ell_{MAX}=0$ PCPA 
calculations for a bcc supercell containing 108 atoms at lattice costant $a=6.88$ a.u.. 
Circles, squares, triangles and crosses stand for Cu, Ni, Zn and Al atoms. 
}
\end{figure}

It is interesting to observe that similar linear 'qV' laws can be derived 
starting from the GCPA functional, Eq.~(\ref{gpotcg}), for random alloys 
by a second order series expansion about the zero Madelung field multipole 
moments, $\{q^0\}$, that can be obtained by solving the following set of 
equations: 
\begin{equation}
\label{q0ldef}
\frac{\partial \widetilde{\omega}^{GCPA}_i}{\partial q_{i,L}} =0
\end{equation}
This procedure leads to a Ginzburg-Landau configurational 'Hamiltonian'
in which the relevant fields are constituted by the values of the 
multipole moments of each lattice site. In formulae:
\begin{widetext}
\begin{equation}
\label{gpotcef}
\widetilde{\Omega}^{CEF}(\{q\}, \mu)=
\frac{1}{2}\sum_{i,L,L^\prime}
w_i a_{i,LL^\prime} (q_{i,L}-q^0_{i,L}) (q_{i,L^\prime}-q^0_{i,L^\prime})
+\frac{1}{2} \sum_{i,j,L,L^\prime} w_i w_j  
\lambda_L  \lambda_{L^\prime}  
M_{ij,LL^\prime} q_{i,L} q_{j,L^\prime} -
\mu \sum_i q_{i,00}
\end{equation}
\end{widetext}

where we have omitted the term $\widetilde{\Omega}^{CEF}(\{q^0\}, \mu=0)$
that represent the GCPA energy at zero Madelung field and chemical potential
and that is constant in a fixed concentration ensemble. The coefficients
$a_{i,LL^\prime}$ are given by the second derivatives of the GCPA 
functional
\begin{equation}
\label{alldef}
a_{i,LL^\prime}= \bigg( \frac{\partial^2 \widetilde{\omega}^{GCPA}_i(\{q_i\})}
{\partial q_{i,L} \partial q_{i,L^\prime}} \bigg)_{\{q_i\}=\{q^0_i\}} 
\end{equation}

The functional of Eq.~(\ref{gpotcef}) constitutes a generalization
of the Charge Excess Function (CEF) proposed in 
Ref.~\onlinecite{CEF} for discussing the charge transfers in 
metallic alloys and shall then be referred in the following to as the 
CEF. The novel feature here is that Eq.~(\ref{gpotcef}) includes
not only the charge excesses, $q_{i,00}$, but also the charge multipole 
moments with $\ell>0$. 

The minimization of the CEF functional $\widetilde{\Omega}^{CEF}$
wrt. its variables, the set of the multipole moments, $\{q\}$, and
the chemical potential, $\mu$, gives 
\begin{equation}
\label{gleq}
\sum_{L^\prime}
\big[ a_{i,LL^\prime} (q_{i,L^\prime}-q^0_{i,L^\prime}) + 
M_{LL^\prime,ij} \; q_{j,L^\prime}\big] = \mu \delta_{i,00}
\end{equation}
and
\begin{equation}
\label{electroneut}
\sum_i q_{i,00}=0
\end{equation}

Using the definition of the Madelung potentials, Eq.~(\ref{vmadlam}), 
and setting 
\begin{equation}
\label{kdef}
k_{i,L}=\sum_{L^\prime} a_{i,LL^\prime} q^0_{i,L^\prime} + \mu \delta_{i,00}
\end{equation}
it is easy to show that Eq.~(\ref{gleq}) for $L=(0,0)$ coincides with the 
linear laws given by Eq.~(\ref{qv}). Versions of Eqs.~(\ref{gleq}), 
(\ref{electroneut}) and (\ref{kdef})  
with the angular momentum summations truncated at $\ell=0$
can be found in Ref.~\onlinecite{CEF}.

We wish to highlight that the CEF derivation from the GCPA functional is 
based on the assumption, common to all Ginzburg-Landau 
theories~\cite{Khachaturyan,S2}, that the homogeneously disordered phase,
in the present case the random alloy phase, can the starting point for
a perturbative treatment of ordering or segregation phenomena. As 
discussed in the previous subsection, in the
GCPA context, this amounts to conjecture that the coherent scattering-path
matrix $\underline{\tau}^c$ of a random alloy can be used for obtaining
a physical picture of concentration fluctuations, or, in other words, 
that it is able to {\it represent} such fluctuations. 
 
\begin{figure}
\includegraphics[width=7cm]{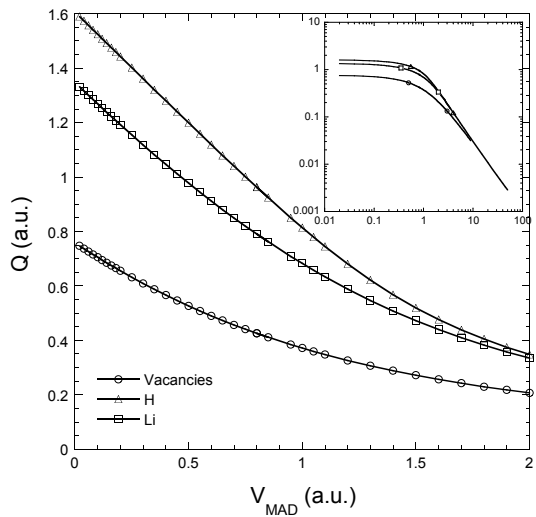}
\caption{\label{lightqv}'qV' relationships for light substitutional 
impurities (vacancies, H and Li atoms) dissolved in bcc Al, from 
CPA+LF calculations. The quantity $Q$ is the number of valence electrons
at the impurity site, therefore $Q=0$ corresponds to 0 electrons for
vacancies and H atoms and 2 electrons for Li. The
linear behaviors observed for small fields ($V_{MAD}< \;1\;a.u.$) are 
superseded by power law trends (see the log-log plot in the inset) for 
very high fields.}
\end{figure}

We wish to close this Section with a few comments. The principal result 
of this paragraph, the CEF functional of Eq.~(\ref{gpotcef}), has been
obtained by a series expansion of the GCPA functional about 
the values that the multipole moments would have in absence of coupling. 
The series has been terminated at the lowest order 
at which differences with respect to SS approximations are expected for.
This, not surprisingly, is enough to obtain a physical picture of the 
charge transfers in metallic alloys. For a given alloy configuration, 
the linear Euler-Lagrange equations obtained by
minimizing the CEF can be easily solved for
the charge multipole moments. The procedure
require the inversion of the matrix $\underline{\underline{F}}$ of
elements
\begin{equation}
\label{fmatrix}
F_{ij,LL^\prime} = a_{i,LL^\prime} \; \delta_{ij} + M_{ij,LL^\prime} 
\end{equation} 

As we have 
shown elsewere~\cite{CEF,brunomatsci}, for a given alloy configuration,
the value of the CEF functional at its minimum has the physical meaning 
of the total energy of the same configuration. The ambiguity due to
the presence of the above mentioned concentration dependent constant 
can be resolved by
comparing CEF and GCPA calculations for a single configuration in a
fixed concentration ensemble. 

In the previous subsection we have described
a general procedure based on the numerical integration of the 'qV' laws
for evaluating the functional form of $\tilde{\omega}_i^{GCPA}(\{q_i\})$.
Of course, if the random alloy $\underline{\tau}^c$ was able to
represent concentration fluctuations and the 'qV' laws were linear, 
the GCPA and the CEF functionals
would be coincident. We do not think that the 'qV' laws can be truly
linear. The argument is as follows. The local excesses of electrons, 
$q_{i,00}$, accordingly with the physical intuition and with the results
plotted in Figs. 5 and 6, are non-increasing functions of the Madelung
potential $V^{MAD}_{i,00}$. If the 'qV' laws were really linear, $q_{i,00}$
would decrease indefinitively and eventually reach unphysical values,
$q_{i,00}<Z_i$ corresponding to negative charge densities. Actually
we expect that the linear laws cannot be any longer valid when all valence 
electrons are expelled from the site. This circumstance would correspond 
to some critical value for the charge excess, say $q_{i,00}^{crit}$. 
Before this critical value is reached, the 'qV' laws should exhibit a 
crossover to an 
asymptotic behavior, say $q_{i,00} \rightarrow q_{i,00}^{crit}$
as $V^{MAD}_{i,00} \rightarrow \infty$. We have tested this conjecture 
by executing CPA+LF calculations~\cite{CPALF} for a single impurities,
vacancies, H or Li atoms, embedded in Al. The results are shown in 
Fig. 7, where we plot $Q=q_{i,00}-q_{i,00}^{crit}$ for the impurity site
as a function of the relevant Madelung fields. In all the
cases considered, a linear regime is clearly visible at low fields.
A very high fields, a crossover to a power law dependence is observed,
with the number of electrons tending to the critical value from above.
The crossover field is comparable with the host band width. We recall
that in CPA+LF calculations $\underline{\tau}^c$ is that of the host
while the Madelung potential is just an adjustable papameter. While
this is a sensible way for studying the response of the impurity
to the perturbing field, this do not imply that all the range of
the perturbations considered is physically meaningful. We do not think
that such high fields, corresponding to the tunneling regime in
the impurity site, could occur in real systems as this would require
a too large defect of electrons at the impurity nearest neighbors. 
Hence, Fig. 7, while supporting the view that the linearity of the 'qV' 
laws and the CEF
are just approximations, does not support the possibility that,
at least for metallic systems, appreciable deviations from linearity or
failures of the CEF are likely to occur.

Another point we wish to address is concerned with the value of the 
chemical potential $\mu$ in Eq.~(\ref{gpotcef}). In a recent paper,
Drchal et al.~\cite{Drchal} argued that $\mu$ should be always zero since
the Fourier transform of the Madelung coefficients
with $L=L^\prime=(0,0)$ diverges as $k \rightarrow \infty$ implying
that the sum of the charge excesses $\sum q_{i,00}$ must vanish, 
automatically satisfying the electroneutrality constraint. The
observation of Drchal et al. is correct for infinite systems, while
for finite supercells, even with periodic boundary condition, the
same Fourier transform always remains finite. $k$, in fact can take 
only the values of the reciprocal space vectors that consitute the
tiling of the supercell considered~\cite{NLCPAII}. The set of the 
allowed values for $k$ includes $0$ only for infinitely large 
supercells. In most practical calculations, then, $\mu$ is necessary, 
although usually it takes small non-zero values.

\section{Numerical results}
In this Section we present a series of numerical tests designed to
study the limits 
of validity of the GCPA and CEF theoretical frameworks. The central issues 
here shall be investigating the realm of validity of the linear 'qV' laws, 
Eq.~(\ref{qv}) or~(\ref{gleq}), and of the energetics implied by the CEF 
functional, Eq.~(\ref{gpotcef}). Furthermore, we shall try to answer two 
questions:
(i) to what an extent the CEF is able to approximate GCPA calculations and
(ii) how do the predictions from the CEF and the GCPA compare vs. 'exact'
DFT calculations for ordered systems. The GCPA theory chosen for these 
tests is the PCPA~\cite{Ujfalussy},
that, being based on a supercell approach, allows for easy comparison vs.
'exact' DFT calculations.  

Several kinds of calculations shall be presented in this Section. The 'exact' 
DTF results used for comparison shall be LDA full-potential 
LAPW calculations produced using the WIEN2K {\it ab initio} 
package~\cite{WIEN2Ka,WIEN2Kb}. They are referred below 
to as LAPW. In all the cases about 10$^4$ k-points in the full Brillouin 
zone have been used, the spherical harmonics expansion of the potentials
in the muffin-tin spheres has been truncated at $\ell=6$  and the parameter 
$R_{MT} \cdot K_{MAX}$ has been set to 7. 
The PCPA calculations have been performed by a conveniently modified version of
our KKR-CPA code~\cite{numerics}. All PCPA calculations are based on the 
ASA approximation for the site potentials, use several thousands k-points 
in the full Brillouin Zone and 31 energies over a complex integration 
contour. For both LAPW and PCPA calculations, the 
core electrons treatment is fully relativistic while a non-relativistic 
approximation is used for valence states. Finally, we present CEF 
calculations~\cite{CEF,philmag} with the charge multipolar expansion 
truncated at $\ell=0$. The concentration dependent parameters 
required by the CEF have been 
obtained from the linear regressions of the 'qV' data generated from 
supercells with random occupancies and the required mean atomic 
concentrations and are reported in Tables~\ref{tabi} and ~\ref{tabii}. 
Depending on which was the source of the parameters, the
CEF calculations shall be referred to as CEF-PCPA or CEF-LAPW. Using the 
formalism of the previous Section, for both PCPA and CEF calculations we
set $w_i=1$ for all lattice sites, $\lambda_{00}=1$ and 
$\lambda_{\ell m}=0$ for $\ell >0$.

\begin{table}
\caption{\label{tabi} CEF parameters obtained by the linear regression of
the 'qV' data from PCPA calculations for random Cu$_c$Zn$_{1-c}$
alloys in bcc or fcc lattices.
$C_{Cu}$ and $C_{Zn}$ are defined as the difference between 1 and the 
correlations obtained from the regressions for the Cu and Zn site charges.
All the quantities are expressed in atomic units. CEF calculations
using the coefficients presented in this Table are referred to as
CEF-PCPA. 
}
\begin{ruledtabular}
\begin{tabular}{cc|ccccc}
  &  c  & $a_{Cu}$ & $a_{Zn}$ & $k_{Cu}-k_{Zn}$ & $C_{Cu}$ & $C_{Zn}$ \\
\hline
    &   0.20 & 1.223 & 1.211 & 0.146 & 3~10$^{-7}$ & 1~10$^{-6}$    \\
    &   0.25 & 1.225 & 1.214 & 0.147 & 4~10$^{-7}$ & 1~10$^{-6}$    \\
    &   0.33 & 1.223 & 1.215 & 0.148 & 5~10$^{-7}$ & 2~10$^{-6}$    \\
bcc &   0.50 & 1.219 & 1.214 & 0.146 & 5~10$^{-7}$ & 2~10$^{-6}$    \\
    &   0.67 & 1.215 & 1.214 & 0.144 & 3~10$^{-7}$ & 2~10$^{-6}$    \\
    &   0.75 & 1.214 & 1.214 & 0.144 & 3~10$^{-7}$ & 2~10$^{-6}$    \\
    &   0.80 & 1.213 & 1.214 & 0.144 & 3~10$^{-7}$ & 9~10$^{-7}$    \\
\hline
    &   0.20 & 1.220 & 1.212 & 0.138 & 3~10$^{-7}$ & 1~10$^{-6}$    \\
    &   0.25 & 1.221 & 1.214 & 0.140 & 3~10$^{-7}$ & 9~10$^{-7}$    \\
    &   0.33 & 1.222 & 1.216 & 0.142 & 3~10$^{-7}$ & 1~10$^{-6}$    \\
fcc &   0.50 & 1.222 & 1.217 & 0.143 & 5~10$^{-7}$ & 2~10$^{-6}$    \\
    &   0.67 & 1.223 & 1.222 & 0.145 & 5~10$^{-7}$ & 1~10$^{-6}$    \\
    &   0.75 & 1.222 & 1.223 & 0.145 & 3~10$^{-7}$ & 1~10$^{-6}$    \\
    &   0.80 & 1.222 & 1.222 & 0.145 & 3~10$^{-7}$ & 2~10$^{-6}$    \\

\end{tabular}
\end{ruledtabular}
\end{table}

\begin{table}
\caption{\label{tabii} CEF parameters obtained by the linear regression of
the 'qV' data from LAPW calculations for random Cu$_c$Zn$_{1-c}$
alloys in bcc or fcc lattices.
$C_{Cu}$ and $C_{Zn}$ are defined as the difference between 1 and the 
correlations obtained from the regressions for the Cu and Zn site charges.
All the quantities are expressed in atomic units. CEF calculations
using the coefficients presented in this Table are referred to as
CEF-LAPW.
}
\begin{ruledtabular}
\begin{tabular}{cc|ccccc}
  &  c  & $a_{Cu}$ & $a_{Zn}$ & $k_{Cu}-k_{Zn}$ & $C_{Cu}$ & $C_{Zn}$ \\
\hline
    &   0.25 & 2.968 & 2.181 & 0.456 & 3~10$^{-2}$ & 2~10$^{-2}$    \\
    &   0.33 & 2.704 & 2.327 & 0.445 & 4~10$^{-3}$ & 8~10$^{-3}$    \\
bcc &   0.50 & 2.811 & 2.307 & 0.413 & 9~10$^{-3}$ & 5~10$^{-2}$    \\
    &   0.67 & 3.388 & 3.351 & 0.590 & 3~10$^{-3}$ & 7~10$^{-3}$    \\
    &   0.75 & 2.586 & 2.652 & 0.432 & 3~10$^{-2}$ & 1~10$^{-2}$    \\
\hline
    &   0.25 & 2.457 & 1.949 & 0.360 & 5~10$^{-3}$ & 9~10$^{-4}$    \\
fcc &   0.50 & 2.287 & 2.130 & 0.350 & 9~10$^{-3}$ & 3~10$^{-3}$    \\
    &   0.75 & 2.646 & 2.317 & 0.399 & 4~10$^{-3}$ & 2~10$^{-4}$    \\

\end{tabular}
\end{ruledtabular}
\end{table}

\subsection{qV laws}
In Sect. III we have presented the CEF functional as an approximation for 
the GCPA functional and have shown how this is equivalent to assume the 
linearity of the 'qV' laws. In Figs. (5) and (6), we plot the 'qV' 
curves from our PCPA calculations for the binary bcc 
Cu$_{0.50}$Zn$_{0.50}$ and the quaternary 
Al$_{0.25}$Cu$_{0.25}$Ni$_{0.25}$Zn$_{0.25}$ fcc random alloys.
It is surprising to observe how much accurately the PCPA data can be
fitted by straight lines. The correlations coefficients obtained from the 
linear regression of the same data differ from unit by about $10^{-6}$.
Similar very high correlations are always obtained from the analysis of PCPA
'qV' data, as it is evident by looking at Table~\ref{tabi}. As it
is shown in Table~\ref{tabii}, also LAPW data present high 
correlations, although the corresponding linear fits are not perfect
and their correlations deviate from unit by $10^{-2}$ or $10^{-3}$.
This notwithstanding, as argued in Sect. III C, we believe that the 
linearity of the 'qV' relationships within the PCPA is just an 
approximation. In order to
check out how much accurate it is, we have studied one of 
the most difficult realistic cases, that of a high charge transfer
ordered alloy, namely the CuZn system. This system has been
studied with many different theoretical 
approaches~\cite{Althoff,Crisan,alphabrass,cunizn,optcuzn,Tulip}. 
It is also relevant, for our present
concerns, that the total energy differences
between fcc and bcc geometrical alloy 
arrangements are relatively small. We have executed calculations
for all the set of 62 bcc and fcc based structures reported in 
Refs.~\onlinecite{Curtarolo} and~\onlinecite{Curtarolothesis}. These 
structures include several ordered crystals for each of the 
the following Cu atomic concentrations: 
0.20, 0.25, 0.33, 0.50, 0.66, 0.75 and 0.80. 
In order to facilitate the comparison, the lattice
constants have been kept fixed to the values 5.5 and 6.9 a.u., respectively
for bcc and fcc based lattices. The results for bcc- and fcc-based alloys 
are reported in Tables~\ref{tabiii} and~\ref{tabiv}, respectively.  

\begingroup
\squeezetable
\begin{table*}
\caption{\label{tabiii}
\scriptsize{Charge excesses and total energies per atom for bcc-based Cu$_c$Zn$_{1-c}$ 
alloys. The first two columns on the left give, for each system, 
the mean Cu atomic concentration ($c$) and, when available, the 
supercell 
identifier in the database of Ref.~\cite{Curtarolothesis} (conf). 
"R" followed by a number, e.g., R16, stands for a quasirandom supercell 
containing  the corresponding number of atoms not included in the database. 
In the third column $n_{unl}$ indicates the mean number of unlike 
nearest neighbors of Zn sites.
The columns from 4th to 7th report the MSD of the charge excesses, 
$<(\Delta q)^2>$, obtained 
by the comparison of different theories: PCPA vs. 
CEF-PCPA (a), LAPW vs. CEF-PCPA (b), LAPW vs. CEF-LAPW (c), LAPW vs.
the model of Ref.~\onlinecite{Magri} (d). Columns from 8th to 11th:
total energies per atom from PCPA, CEF-PCPA, LAPW and CEF-LAPW
calculations. The energy zero is given, for each concentration, by the
PCPA prediction for the ground state.}}
\begin{ruledtabular}
\begin{tabular}{ccc|cccc|cccc}
       &  &   & \multicolumn{4}{c|}{$<(\Delta q)^2>$} & \multicolumn{4}{c}{$\Delta E$ (mRy)} \\ \cline{4-11}
 c &  conf & $n^{unl}$ & \textbf{a} & \textbf{b} & \textbf{c} & \textbf{d} & PCPA & CEF-PCPA & LAPW & CEF-LAPW \\ \hline

       & 92 & 1.5 & 5~10$^{-11}$ & 1~10$^{-4}$ &    -  &  -  & 0.038 & 0.040 & -0.015 &  -  \\
  0.20 & 98 & 2.0 & 2~10$^{-9}$ & 4~10$^{-4}$ &    -  &  -  & 0.000 & 0.000 & 0.000 &  -  \\
\hline
       & 69 & 2.0 & 3~10$^{-9}$ & 4~10$^{-5}$ &  2~10$^{-6}$ &  -  & 0.741 & 0.742 & 0.991 & 0.556   \\
       & 72 & 1.3 & 9~10$^{-9}$ & 4~10$^{-4}$ &  2~10$^{-5}$ &  -  & 1.966 & 1.971 & 3.951 & 2.437   \\
       & 75 & 2.0 & 1~10$^{-9}$ & 1~10$^{-5}$ &  1~10$^{-6}$ &  -  & 0.824 & 0.823 & 1.089 & 0.631   \\
  0.25 & 78 & 2.7 & 1~10$^{-9}$ & 4~10$^{-4}$ &  5~10$^{-7}$ &  -  & 0.865 & 0.864 & 0.980 & 0.838   \\
       & 81 & 2.7 & 3~10$^{-8}$ & 2~10$^{-4}$ &  4~10$^{-7}$ &  -  & 0.353 & 0.353 & 0.050 & 0.246   \\
       & 83 & 2.7 & 2~10$^{-8}$ & 3~10$^{-4}$ &  1~10$^{-6}$ &  -  & 0.327 & 0.327 & 0.194 & 0.260   \\
       & 86 & 2.7 & 6~10$^{-8}$ & 4~10$^{-4}$ &  2~10$^{-5}$ &  -  & 0.000 & 0.000 & 0.000 & 0.000   \\
       & R16 & 2.2 & 1~10$^{-8}$ & 1~10$^{-4}$ &  3~10$^{-6}$ &  -  & 0.797 & 0.799 & 1.117 & 0.669   \\
\hline
       & 63 & 3.0 & 3~10$^{-8}$ & 2~10$^{-4}$ &  2~10$^{-5}$ &  -  & 0.000 & 0.000 & 0.000 & 0.000   \\
  0.33 & 65 & 2.0 & 2~10$^{-9}$ & 3~10$^{-4}$ &  2~10$^{-5}$ &  -  & 1.741 & 1.747 & 2.950 & 2.129   \\
       & 67 & 4.0 & 2~10$^{-8}$ & 2~10$^{-4}$ &  3~10$^{-5}$ &  -  & 0.078 & 0.078 & -0.519 & 0.068   \\
       & R18 & 3.3 & 6~10$^{-9}$ & 3~10$^{-4}$ &  4~10$^{-5}$ &  -  & 0.558 & 0.562 & 0.729 & 0.838   \\
\hline
       & 60 & 4.0 & 4~10$^{-9}$ & 4~10$^{-5}$ &  5~10$^{-6}$ & 3~10$^{-4}$ & 1.661 & 1.662 & 3.457 & 1.188   \\
       & 61 & 8.0 & 3~10$^{-9}$ & 1~10$^{-3}$ &  6~10$^{-6}$ & 3~10$^{-3}$ & 0.000 & 0.000 & 0.000 & 0.000   \\
       & 71 & 2.0 & 2~10$^{-9}$ & 1~10$^{-3}$ &  4~10$^{-5}$ & 9~10$^{-4}$ & 4.107 & 4.115 & 8.657 & 5.089   \\
  0.50 & 74 & 4.0 & 3~10$^{-9}$ & 5~10$^{-7}$ &  4~10$^{-7}$ & 3~10$^{-4}$ & 1.823 & 1.824 & 3.666 & 1.342   \\
       & 77 & 4.0 & 4~10$^{-11}$ & 2~10$^{-4}$ &  1~10$^{-6}$ & 7~10$^{-5}$ & 2.736 & 2.739 & 4.804 & 2.404   \\
       & 80 & 6.0 & 7~10$^{-9}$ & 4~10$^{-4}$ &  8~10$^{-7}$ & 3~10$^{-4}$ & 0.885 & 0.883 & 1.666 & 0.557   \\
       & 85 & 4.0 & 7~10$^{-9}$ & 2~10$^{-4}$ &  1~10$^{-5}$ & 7~10$^{-4}$ & 1.007 & 1.006 & 2.757 & 0.646   \\
       & R16 & 4.3 & 3~10$^{-9}$ & 2~10$^{-4}$ &  3~10$^{-6}$ & 4~10$^{-4}$ & 1.989 & 1.989 & 3.806 & 1.613   \\
\hline
       & 62 & 6.0 & 7~10$^{-10}$ & 2~10$^{-4}$ &  6~10$^{-7}$ &  -  & 0.000 & 0.000 & 0.000 & 0.000   \\
  0.67 & 64 & 4.0 & 8~10$^{-12}$ & 3~10$^{-4}$ &  2~10$^{-5}$ &  -  & 1.698 & 1.703 & 2.671 & 2.021   \\
       & 66 & 8.0 & 2~10$^{-10}$ & 1~10$^{-4}$ &  2~10$^{-7}$ &  -  & 0.076 & 0.077 & -0.650 & 0.061   \\
       & R18 & 6.7 & 3~10$^{-9}$ & 3~10$^{-4}$ &  1~10$^{-6}$ &  -  & 0.545 & 0.549 & 0.508 & 0.870   \\
\hline
       & 68 & 6.0 & 2~10$^{-11}$ & 2~10$^{-5}$ &  1~10$^{-6}$ &  -  & 0.726 & 0.732 & 1.364 & 0.566   \\
       & 70 & 4.0 & 3~10$^{-9}$ & 6~10$^{-4}$ &  3~10$^{-5}$ &  -  & 1.927 & 1.935 & 3.395 & 2.505   \\
       & 73 & 6.0 & 2~10$^{-10}$ & 9~10$^{-6}$ &  2~10$^{-6}$ &  -  & 0.806 & 0.812 & 1.390 & 0.642   \\
  0.75 & 76 & 8.0 & 1~10$^{-9}$ & 5~10$^{-4}$ &  4~10$^{-6}$ &  -  & 0.850 & 0.852 & 1.022 & 0.873   \\
       & 79 & 8.0 & 3~10$^{-9}$ & 2~10$^{-4}$ &  1~10$^{-6}$ &  -  & 0.345 & 0.349 & 0.464 & 0.251   \\
       & 82 & 8.0 & 4~10$^{-9}$ & 3~10$^{-4}$ &  2~10$^{-6}$ &  -  & 0.322 & 0.323 & 0.462 & 0.269   \\
       & 84 & 8.0 & 3~10$^{-8}$ & 4~10$^{-4}$ &  4~10$^{-6}$ &  -  & 0.000 & 0.000 & 0.000 & 0.000   \\
       & R16 & 6.8 & 1~10$^{-9}$ & 2~10$^{-4}$ &  3~10$^{-6}$ &  -  & 0.782 & 0.788 & 1.278 & 0.685   \\
\hline
       & 87 & 6.0 & 5~10$^{-10}$ & 1~10$^{-4}$ &    -  &  -  & 0.033 & 0.038 & 0.327 &  -  \\
  0.80 & 93 & 8.0 & 4~10$^{-10}$ & 6~10$^{-4}$ &    -  &  -  & 0.000 & 0.000 & 0.000 &  -  \\
\end{tabular}
\end{ruledtabular}
\end{table*}
\endgroup

The charges from the CEF-PCPA are, in practice, identical to
those obtained from the
PCPA theory for the ordered systems. In order to represent the size 
of these tiny differences, we report in Tables~\ref{tabiii} and~\ref{tabiv}
the mean square 
displacement between the two sets of calculated charges, $<(\Delta q)^2>$. 
In the worst case, the bcc-based DO$_2$ structure, identified in 
Table~\ref{tabiv}
by the number 86, we find $<(\Delta q)^2>=6~10^{-8}$.
Such an excellent agreement has been obtained for all the set of ordered 
structures
considered, in spite of the fact that the CEF input has been obtained 
from random supercells.

\begingroup
\squeezetable
\begin{table*}
\caption{\label{tabiv}
\scriptsize{Charge excesses and total energies per atom for fcc-based Cu$_c$Zn$_{1-c}$ 
alloys. The first two columns on the left give, for each system, 
the mean Cu atomic concentration ($c$) and, when available, the 
supercell 
identifier in the database of Ref.~\cite{Curtarolothesis} (conf). 
"R" followed by a number, e.g., R16, stands for a quasirandom supercell 
containing  the corresponding number of atoms not included in the database. 
In the third column $n_{unl}$ indicates the mean number of unlike 
nearest neighbors of Zn sites.
The columns from 4th to 7th report the MSD of the charge excesses, 
$<(\Delta q)^2>$, obtained 
by the comparison of different theories: PCPA vs. 
CEF-PCPA (a), LAPW vs. CEF-PCPA (b), LAPW vs. CEF-LAPW (c), LAPW vs.
the model of Ref.~\onlinecite{Magri} (d). Columns from 8th to 11th:
total energies per atom from PCPA, CEF-PCPA, LAPW and CEF-LAPW
calculations. The energy zero is given, for each concentration, by the
PCPA prediction for the ground state.}}
\begin{ruledtabular}
\begin{tabular}{ccc|cccc|cccc}
       &  &   & \multicolumn{4}{c|}{$<(\Delta q)^2>$} & \multicolumn{4}{c}{$\Delta E$ (mRy)} \\ \cline{4-11}
 c &  conf & $n^{unl}$ & \textbf{a} & \textbf{b} & \textbf{c} & \textbf{d} & PCPA & CEF-PCPA & LAPW & CEF-LAPW \\ \hline

       & 35 & 2.5 & 3~10$^{-10}$ & 2~10$^{-5}$ &    -  &  -  & 0.500 & 0.503 & 0.926 &  -  \\
  0.20 & 39 & 3.0 & 7~10$^{-9}$ & 7~10$^{-5}$ &    -  &  -  & 0.000 & 0.000 & 0.000 &  -  \\
\hline
       & 12 & 3.3 & 2~10$^{-9}$ & 7~10$^{-5}$ &  4~10$^{-6}$ &  -  & 0.515 & 0.516 & 1.078 & 0.438   \\
       & 15 & 2.7 & 1~10$^{-8}$ & 2~10$^{-4}$ &  4~10$^{-7}$ &  -  & 1.379 & 1.386 & 2.151 & 1.658   \\
       & 18 & 3.3 & 2~10$^{-9}$ & 2~10$^{-5}$ &  2~10$^{-6}$ &  -  & 0.566 & 0.568 & 1.054 & 0.471   \\
  0.25 & 21 & 3.3 & 2~10$^{-9}$ & 1~10$^{-4}$ &  8~10$^{-6}$ &  -  & 0.680 & 0.685 & 1.510 & 0.616   \\
       & 24 & 4.0 & 2~10$^{-8}$ & 2~10$^{-4}$ &  5~10$^{-6}$ &  -  & 0.000 & 0.000 & 0.000 & 0.000   \\
       & 26 & 4.0 & 2~10$^{-8}$ & 2~10$^{-4}$ &  3~10$^{-7}$ &  -  & 0.068 & 0.069 & 0.073 & 0.051   \\
       & 29 & 2.0 & 5~10$^{-9}$ & 4~10$^{-4}$ &  2~10$^{-5}$ &  -  & 1.817 & 1.824 & 4.202 & 2.349   \\
       & R16 & 3.0 & 1~10$^{-8}$ & 2~10$^{-4}$ &  1~10$^{-6}$ &  -  & 1.015 & 1.020 & 1.705 & 1.169   \\
\hline
       & 6 & 4.0 & 1~10$^{-10}$ & 5~10$^{-5}$ &    -  &  -  & 1.179 & 1.185 & 1.016 &  -  \\
  0.33 & 8 & 5.0 & 1~10$^{-8}$ & 1~10$^{-4}$ &    -  &  -  & 0.000 & 0.000 & 0.000 &  -  \\
       & 10 & 3.0 & 6~10$^{-11}$ & 3~10$^{-4}$ &    -  &  -  & 1.800 & 1.807 & 3.300 &  -  \\
\hline
       & 3 & 8.0 & 7~10$^{-9}$ & 3~10$^{-4}$ &  5~10$^{-6}$ & 1~10$^{-4}$ & 0.139 & 0.144 & -0.075 & 0.111   \\
       & 4 & 6.0 & 2~10$^{-9}$ & 5~10$^{-6}$ &  2~10$^{-5}$ & 6~10$^{-5}$ & 1.075 & 1.081 & 2.141 & 0.961   \\
       & 14 & 4.0 & 1~10$^{-9}$ & 6~10$^{-4}$ &  5~10$^{-6}$ & 2~10$^{-4}$ & 2.895 & 2.905 & 4.803 & 3.657   \\
  0.50 & 17 & 7.0 & 3~10$^{-9}$ & 3~10$^{-5}$ &  2~10$^{-6}$ & 1~10$^{-6}$ & 0.717 & 0.720 & 1.121 & 0.605   \\
       & 20 & 6.0 & 5~10$^{-10}$ & 5~10$^{-5}$ &  1~10$^{-7}$ & 5~10$^{-5}$ & 1.429 & 1.434 & 2.464 & 1.353   \\
       & 23 & 8.0 & 7~10$^{-9}$ & 2~10$^{-4}$ &  1~10$^{-6}$ & 4~10$^{-5}$ & 0.000 & 0.000 & 0.000 & 0.000   \\
       & 28 & 3.0 & 2~10$^{-9}$ & 9~10$^{-4}$ &  6~10$^{-5}$ & 5~10$^{-4}$ & 3.339 & 3.350 & 6.745 & 4.694   \\
       & R16 & 6.8 & 3~10$^{-9}$ & 2~10$^{-4}$ &  2~10$^{-6}$ & 6~10$^{-5}$ & 0.913 & 0.918 & 1.519 & 0.972   \\
\hline
       & 5 & 8.0 & 3~10$^{-10}$ & 9~10$^{-5}$ &    -  &  -  & 1.102 & 1.224 & 1.283 &  -  \\
  0.67 & 7 & 10.0 & 1~10$^{-8}$ & 1~10$^{-4}$ &    -  &  -  & 0.000 & 0.000 & 0.000 &  -  \\
       & 9 & 6.0 & 3~10$^{-9}$ & 3~10$^{-4}$ &    -  &  -  & 1.743 & 1.866 & 3.224 &  -  \\
\hline
       & 11 & 10.0 & 3~10$^{-10}$ & 9~10$^{-5}$ &  2~10$^{-5}$ &  -  & 0.549 & 0.550 & 1.009 & 0.501   \\
       & 13 & 8.0 & 5~10$^{-9}$ & 5~10$^{-4}$ &  3~10$^{-6}$ &  -  & 1.479 & 1.482 & 2.230 & 1.986   \\
       & 16 & 10.0 & 3~10$^{-10}$ & 2~10$^{-5}$ &  1~10$^{-5}$ &  -  & 0.604 & 0.605 & 1.054 & 0.535   \\
  0.75 & 19 & 10.0 & 5~10$^{-10}$ & 1~10$^{-4}$ &  5~10$^{-6}$ &  -  & 0.729 & 0.731 & 1.083 & 0.709   \\
       & 22 & 12.0 & 3~10$^{-9}$ & 2~10$^{-4}$ &  5~10$^{-6}$ &  -  & 0.000 & 0.000 & 0.000 & 0.000   \\
       & 25 & 12.0 & 2~10$^{-9}$ & 2~10$^{-4}$ &  5~10$^{-6}$ &  -  & 0.073 & 0.073 & -0.193 & 0.057   \\
       & 27 & 6.0 & 1~10$^{-8}$ & 5~10$^{-4}$ &  1~10$^{-5}$ &  -  & 1.944 & 1.949 & 3.654 & 2.799   \\
       & R16 & 3.0 & 2~10$^{-9}$ & 3~10$^{-4}$ &  2~10$^{-6}$ &  -  & 1.088 & 1.091 & 1.790 & 1.409   \\
\hline
       & 30 & 10.0 & 1~10$^{-9}$ & 6~10$^{-5}$ &    -  &  -  & 0.544 & 0.550 & -4.023 &  -  \\
  0.80 & 36 & 12.0 & 2~10$^{-8}$ & 7~10$^{-5}$ &    -  &  -  & 0.000 & 0.000 & 0.000 &  -  \\
\end{tabular}
\end{ruledtabular}
\end{table*}
\endgroup
   
In a previous Letter~\cite{CEF} we have shown that the CEF is
able to carefully reproduce the charges from LSMS calculations. Moreover, 
the parameters extracted from ordered structure calculations can be used 
to predict the charges for random structures and vice-versa. The quality
of the CEF predictions was very good either, with $<(\Delta q)^2>$ of the 
order of $10^{-6}$, i.e. about three orders of magnitude less 
than what we have found by the comparison of CEF and PCPA. 
Since the LSMS calculations presented in Ref.~\onlinecite{CEF} were based 
on the ASA, we surmise that the modest lost of accuracy of CEF predictions 
for LSMS wrt. PCPA calculations constitutes a measure of the  
importance of the scattering effects from nearest neighbors. These
effects, in fact, can be accounted for only in a mean field fashion
by the PCPA.   

We have also investigated the effects of the spherical approximation 
for the atomic potentials by executing full-potential LAPW calculations. 
In Fig. 8 we plot the site charge excesses obtained from LAPW vs. the 
number of unlike nearest neighbors of the same sites, for all the 
structures corresponding to equimolar concentrations. In 
Tables~\ref{tabiii} and~\ref{tabiv}
we report the results for $<(\Delta q)^2>$ at all the concentrations. 
As apparent from Fig. 8, the trends of $q_i$ are not easily 
accounted for by the nearest neighbors environment only~\cite{Magri},
expecially for bcc based structures.
\begin{figure}\label{magrifig7}
\includegraphics[width=7cm]{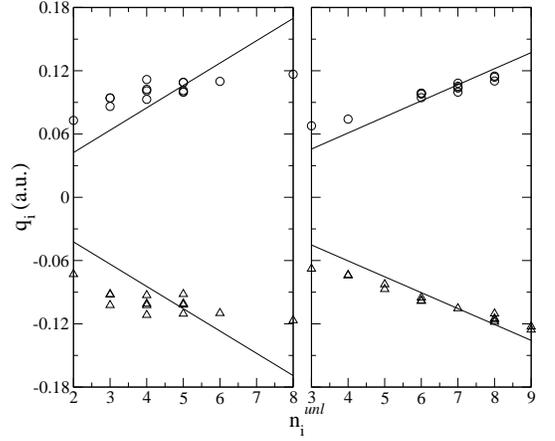}
\caption{
Charge excesses, $q_i$, vs. the number of unlike nearest neighbors of the 
corresponding site, $n^{unl}_i$, from LAPW calculations for many, bcc and fcc 
based, ordered and disordered, 
configurations, of $Cu_{0.50}Zn_{0.50}$ alloys.  
Circles and triangles represent charges on Cu and Zn sites, respectively.
Left frame: bcc-based alloys; right frame: fcc-based alloys. 
The straight lines in each panel indicate the best fits obtained
by the model of Magri et al.~\cite{Magri}. 
}
\end{figure}
This notwithstanding, CEF-PCPA calculations reasonably account for 
the LAPW charges. As it can be 
seen in the columns marked as (b) of Tables~\ref{tabiii} and~\ref{tabiv}, 
$<(\Delta q)^2>$ is usually of 
the order of $10^{-4}$, sometimes less, and about $10^{-3}$ in the
worst case. In order to understand how much these results can be affected  
by the PCPA input coefficients, we have repeated CEF calculations by 
fitting the coefficients from LAPW 'qV' data for the random alloy 
configurations corresponding to the relevant stoichiometries and
reported in Tables~\ref{tabiii} and~\ref{tabiv}. 
As shown in the columns (c) of Tables~\ref{tabiii} and~\ref{tabiv}, this 
reduces $<(\Delta q)^2>$ of about one order of magnitude. 

\begin{figure}\label{dqs8}
\includegraphics[width=7cm]{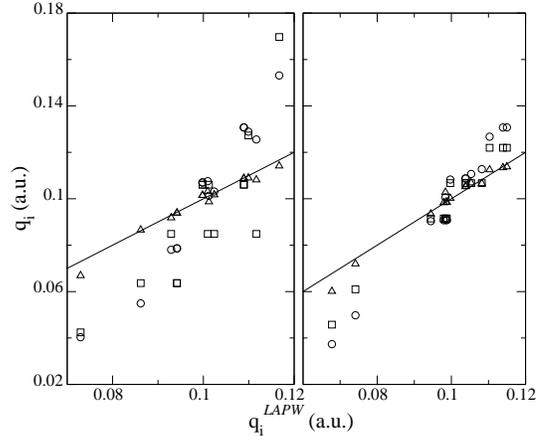}
\caption{
Charge excesses on Cu sites $q_i$ for the same $Cu_{0.50}Zn_{0.50}$ alloys
as in Fig. 8. Circles, triangles and squares represent the calculated 
values by CEF-PCPA, CEF-LAPW and by the model of Ref.~\onlinecite{Magri},
In abscissa the charge excesses obtained by LAPW calculations 
are reported.
Left frame: bcc-based alloys; right frame: fcc-based alloys. 
In oder to improve readability we plot also the straight lines 
$q=q^{LAPW}$. The deviations from these lines measure
the accuracy of the various calculations.
}
\end{figure}

Interestingly, the present CEF-LAPW calculations confirm the 
observations about the transferability of CEF parameters in 
Ref.~\onlinecite{CEF}, where the CEF charges have been compared vs. 
LSMS results. As a typical example, let us consider the results for 
$c=0.50$ 
reported in Table~\ref{tabiii}. The $< \Delta q^2 >$ obtained are alway small:
$4 \; 10^{-5}$ in the worst case and $4\; 10^{-7}$ in the best, corresponding
respectively to structures 71 and 74, while an intermediate value,
$3 \; 10^{-6}$ is found for the structure $R16$, from which the CEF-LAPW coefficients 
have been obtained. The same holds for all the concentrations,
both for bcc and fcc structures. A look to the columns (c) in the 
Tables~\ref{tabiii} and~\ref{tabiv}, in fact, shows that 
while excellent results have been obtained for 
all the supercell, the random structure from which the CEF coefficients have
been extracted is not necessarily the best performing.

The above arguments about the charges should not lead to the conclusion 
that, for this purpose, any fit is comparable 
with any other. This is clearly shown in Fig. 9, where the performances
of PCPA, CEF-PCPA, CEF-LAPW and by the model of Magri et al.~\cite{Magri}
are compared for the equiatomic concentration alloys. 
In the same Figure, the distances from the diagonal 
lines measures the differences between the LAPW charges and those by various 
approximations, for all the Cu sites in the supercells with $c=0.5$. 
It is there evident that the results by CEF-LAPW, marked by open triangles,
are much better than those by other approximations.

\subsection{Total energies}
In Tables~\ref{tabiii} and~\ref{tabiv} we compare the total energies obtained for CuZn 
alloys by CEF, PCPA and LAPW calculations. We have used the same extended 
set of bcc and fcc based structures listed in 
Ref.~\onlinecite{Curtarolothesis}. 
Since the CEF energies contain a, concentration dependent 
constant, we report the quantity
$\Delta E$, defined as the energy difference between the structure
at hand and the structure that, at the same concentration, has the lowest 
energy, according with PCPA calculations. The same $\Delta E$ is plotted 
in Figs. 10 and 11 for the Cu concentrations $c=0.25$, $0.50$, an $0.75$,
for which the database of Ref.~\onlinecite{Curtarolothesis} contains a 
number of
structure sufficient to individuate trends. 

\begin{figure}
\includegraphics[width=7cm]{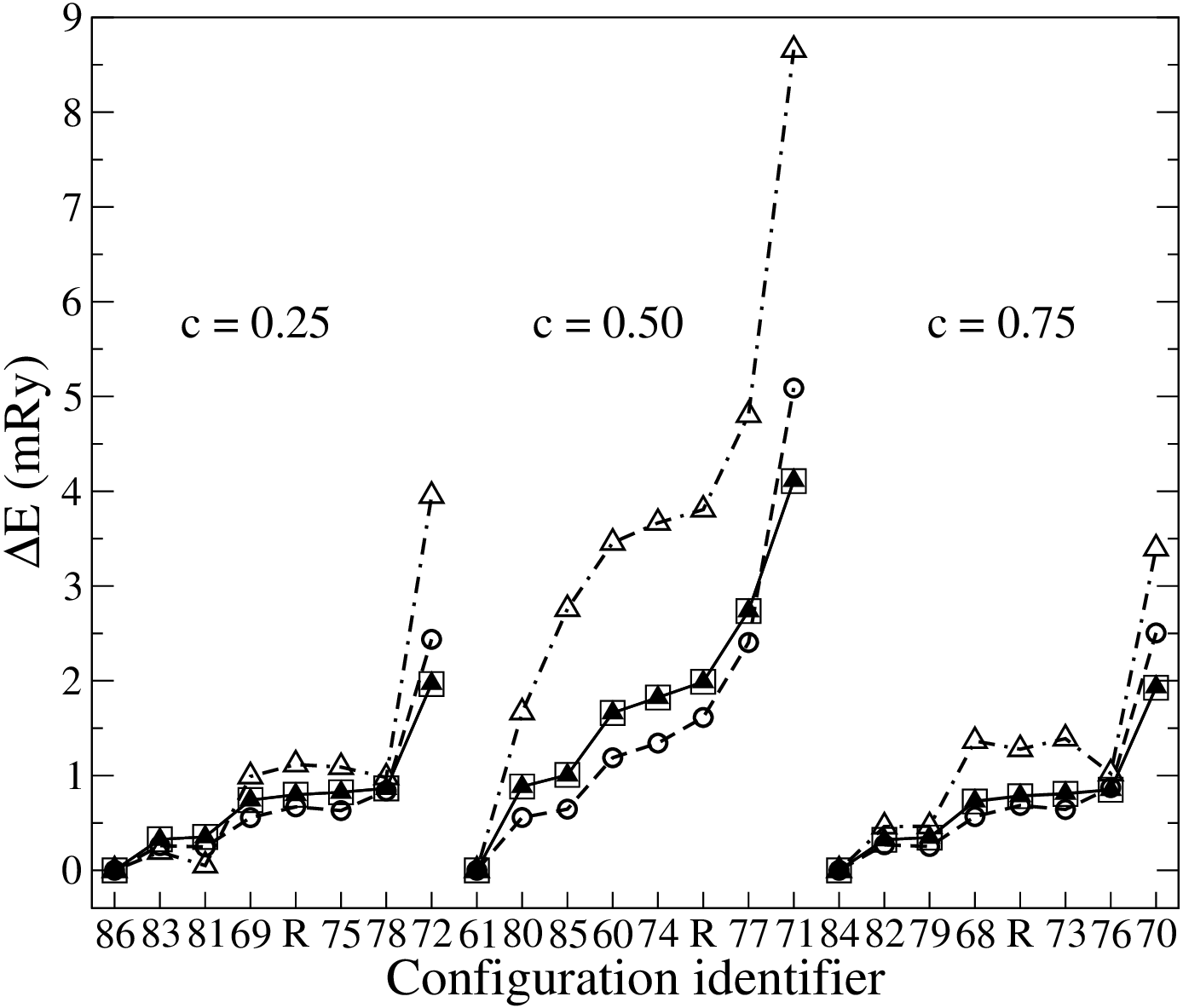}
\caption{\label{fig10} 
Total energy differences with respect to the PCPA
predicted ground state, $\Delta E$, for bcc based
Cu$_c$Zn$_{1-c}$ alloys. The labels indicated in abscissa identify the 
various configurations in the database of Ref.~\onlinecite{Curtarolothesis},  
$R$ stands for structures with randomly generated chemical occupations 
containing 16 atoms with mean  Cu contents c=0.25, 
0.5 and 0.75. Open triangles, open circles,
open squares and filled triangles, indicate LAPW, CEF-LAPW, CEF-PCPA
and PCPA calculations, respectively. Lines are a guide for the eye.
}
\end{figure}

Our first observation is that the total energies obtained by PCPA and 
CEF-PCPA calculations perfectly overlap on the scale of Figs. 10 
and 11, where they are represented as filled triangles and open squares,
respectively.
As reported in Tables~\ref{tabiii} and~\ref{tabiv}, in fact, the values obtained by the two 
methods are different by a few $\mu Ry$ per atom, that is comparable with 
the accuracy of the calculations. Thus, PCPA and CEF-PCPA give indintinguishable 
results both for the charges (as discussed in the previous subsection) and the
total energies. Therefore, it is compelling to conclude that 
the CEF theory is a numerically excellent and powerful tool to reproduce
with much less efforts GCPA electronic structure calculations. Moreover,
since CEF-PCPA calculations use as an input the 'qV'
data obtained from {\it random} supercells, the perfect agreement 
obtained for the properties of so many different {\it ordered} structures 
that have not been used to fit the CEF coefficients has only one possible
explanation. Accordingly with the 
discussion in Sects. III B and III C,
both the following conditions must be fulfilled. (i) The coherent 
scattering-path matrix $\underline{\tau}^c$ of the random alloy 
configuration used as an input must be representative of the whole set
of ordered structures considered; (ii) the linearity of the 'qV' laws
is almost perfectly observed in all the range of values that the charge
excesses and the Madelung potentials take for the structures considered.
In Sect. III we have offered several arguments supporting the validity of
both points above, but we have not been able to provide an analytical
demonstration. We think that the numerical evidence found is
very strong and compelling.  

\begin{figure}
\includegraphics[width=7cm]{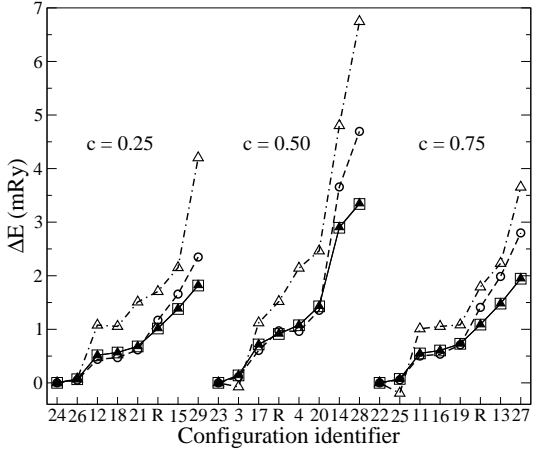}
\caption{\label{fig11} 
Total energy differences with respect to the PCPA
predicted ground state, $\Delta E$, for fcc based
Cu$_c$Zn$_{1-c}$ alloys. The labels indicated in abscissa identify the 
various configurations in the database of Ref.~\onlinecite{Curtarolothesis},  
$R$ stands for structures with randomly generated chemical occupations 
containing 16 atoms with mean  Cu contents c=0.25, 
0.5 and 0.75. Open triangles, open circles,
open squares and filled triangles, indicate LAPW, CEF-LAPW, CEF-PCPA
and PCPA calculations, respectively. Lines are a guide for the eye.
}
\end{figure}

Accordingly with the discussion in Sect. III, the success of the CEF theory in
reproducing the charges or, equivalently, the Madelung potentials guarantees 
the reproducibility of {\it any} ground state 
property within the GCPA theory through Eq.~(\ref{Pii}). Hence, even spectral properties as the 
DOS or the Bloch Spectral functions, although buried, are contained
in the CEF functional that, if the input parameters are extracted
from a GCPA theory, inherits all the good and the bad
things of same GCPA theory.  

The comparison with LAPW calculations is more difficult, for two
different reasons. In first place, these calculations do not 
assume mean boundary conditions for the wave-functions and use a 
procedure equivalent to the full calculation of the 
$\underline{\underline{\tau}}$ matrix. In second place, within LAPW 
calculations, the charge multipole summation is truncated at some high 
$\ell$ value. With these clarifications, the agreement
between LAPW and PCPA or CEF-PCPA calculations (there is no reason 
for discussing the last two models separately) is quite good. 
As a general rule, the two 
set of calculations find the same ground states at the concentrations
considered. In the few
exceptions (that correspond to the negative figures in the LAPW
columns of Tables~\ref{tabiii} and~\ref{tabiv}) the disagreement can be explained by
the fact that the structures indicated as the ground state by the 
two theories are almost degenerate in energy. 
Also the general trends for the total energies are well 
reproduced, as it is visible in Figs. 10 and 11, though the PCPA  
generally underestimate the energy differences. 

Fitting the CEF parameters from the LAPW 'qV' laws generally improves
the agreement. At variance of what found for the charges,
however, the improvement is quite modest.  

In summary: the CEF appears able to perfectly reproduce GCPA calculations
for both ordered and disordered metallic systems. The reasons why the
agreement is so excellent are not yet completely understood. Although 
CEF and GCPA theories are both coarse grained versions of the DFT, 
opposite to what numerical results suggest, they are not the 
{\it same} theory. 
In fact, as discussed in Sect. III, 
in order to be coincident to the CEF, GCPA theories should 
(i) exactly observe linear 'qV' laws and (ii) lead to coherent scattering
matrices independent on the configuration in a fixed concentration ensemble.
For metallic alloys, these conditions appear plausible and the numerical 
evidence strongly support the view that both are nearly satisfied. However we 
must highlight that the condition (i) is not verified for pathologically
high values of the Madelung field. The comparison vs. LAPW calculations
suggest that both coarse grained theories, GCPA and CEF, are able to reproduce
semiquantitatively the total energies of the alloy configurations considered.
In particular, the results by the coarse grained theories are strongly
correlated with those by LAPW. This fact is better elucidated by Figs. 10 
and 11, where configurations belonging 
to the same fixed concentration ensemble are ordered in such a way to have
increasing PCPA total energies. If the same ordering was not observed by
some other method for some configuration, this would show up as a local
minimum in the corresponding curve. The most visible of such events, 
occurs in Fig. 11 for $c=0.75$, where the curve corresponding to LAPW 
calculations
presents a very weak local minimum at the configuration 25. The 
examination of Figs. 10 and 11 suggests that the
coarse grained theories are able to give qualitatively correct predictions
about ordering for the alloys considered, while the fact that they generally
underestimate the corresponding energies could imply incorrect estimates of
the corresponding transition temperatures.

\section{Conclusions} 
We wish to conclude this paper with a summary and a few comments.

We have introduced the class of the GCPA theories, that are characterized
by (i) a specific ansatz for the kinetic part of the density functional, 
which is common to all CPA-based theories, and (ii) an external model that
determines the way in which the atomic effective potentials should be
reconstructed and the statistical weights to be assigned each. The GCPA
class of approximations includes most existing CPA-based density
functional theories, to mention a few: the CPA prototype, i.e. the single 
site CPA~\cite{DFTKKRCPA1,DFTKKRCPA2}, the Screened Impurity Model CPA 
(SIM-CPA)~\cite{SIMCPAI,SIMCPAII}, the Polymorphous CPA 
(PCPA)~\cite{Ujfalussy}, the CPA including Local Fields 
(CPA+LF)~\cite{CPALF}, the Non Local CPA (NL-CPA)~\cite{NLCPAI}. The 
ansatz (i) consists in applying averaged boundary conditions at the 
surfaces of each scattering volume and naturally leads to algorithms 
requiring
a number of operation that scales as $N$. As it is discussed by Abrikosov 
and Johansson~\cite{Abrikosov_cpa}, CPA-based approximations allow 
for a careful picture
of the spectral properties of metallic alloys. The so much criticized
results of the SS-CPA about the total alloy energies can be healed by
external models that consider the charge distribution in the system. We 
have shown how this can be done systematically by writing the relevant 
energetic contributions as a series involving the charge multipole 
moments in each scattering volume. The truncation errors of the same 
series are probably already quite small when only the first term is 
included, as in the case of spherical approximations.

We have derived an expression of the GCPA density functional that,
together with the above multipole sums, 
includes local 'atomic' terms, completely determined by the atomic number
of the ion in the volume, and by the geometry of the same volume. 
The local term at the $i$-th site is coupled to the others only through
the coherent scattering matrix $\underline{\tau}^c$ and the Madelung 
potential at the same site. Although this kind of coupling, that 
we have called {\it marginal coupling}, is not
necessarily weak, nevertheless, it is analytically 
tractable and it is the source of the $O(N)$ scaling in GCPA theories.
We have demonstrated that in a GCPA theory all ground state 
properties within a specific sample are functions of the appropriate 
coupling Madelung potential
{\it only}, or, equivalently, of the charge multipole moments at each 
lattice site. To put it into other words: we have demonstrated that 
the GCPA approximations realize a {\it coarse graining} of the 
Hohenberg-Kohn density functional, since only a part of the 
information conveyed by the electronic density field, namely the
charge multipole moments, is actually entering in the GCPA approximate
functional. Moreover we have suggested that the explicit form of the
GCPA functional dependence on the multipole moments can be obtained 
in a fixed concentration ensemble by
the numerical integration of the 'qV' relationships for a random alloy 
configuration belonging to the same ensemble. The above procedure does
not rely on the linearity of the 'qV' laws.

We have re-derived the CEF~\cite{CEF} as a sensible approximation of the 
GCPA theories, with which it would coincide provided the 'qV' 
were exactly linear as claimed by many groups. The present derivation 
allows for the inclusion of higher order multipole moments. 
A very remarkable feature of the CEF theory is that it 
shares the same structure of the MST. In fact,
the minimization of the CEF requires the solution of
a set of Euler-Lagrange equations that has the same 
structure of the Korringa-Kohn-Rostoker (KKR) matrix at zero energy 
and wave-vector~\cite{brunomatsci,Drchal}.
More specifically, as it can be seen by comparing Eqs.~(\ref{MKKR})
and~(\ref{fmatrix}), the site-diagonal response functions, 
$a_{i,LL^\prime}$,
and the Madelung coefficients, $M_{ij,LL^\prime}$, in the CEF theory, play
the role of the site diagonal scattering matrices and the KKR structure
constants in the MST theory. The correspondence is not only formal, since 
the $a_{i,LL^\prime}$ are single-site quantities in the same sense of 
the SS scattering matrices~\cite{brunomatsci} and, in plain analogy with
them, are related with the SS response to the appropriate perturbing
field~\cite{CPALF}. 

In the present paper we have provided several formal arguments and strong
numerical evidences that CEF and GCPA theories lead to very similar 
results, the discrepancies being of the order of the numerical errors.
We have also shown that CEF and GCPA theories are able to reproduce the
charges and the total energies for many ordered alloy configurations.
In our view the {\it coarse grained theories}, GCPA and CEF, constitute
a valuable alternative to full DFT calculations.

Although the CPA theory was proposed many year ago with the purpose 
of dealing with substitutionally disordered alloys, we think that we 
have shown that today GCPA theories are able to deal with ordered
intermetallic compounds too. Therefore, the fact that CPA-based theories are able to 
cope with sophisticated model of disorder is not an original sin but,
rather, an added value.

The computational performances of the CEF have been discussed in more
details elsewhere~\cite{CEF}. Here we like to mention that the possibility
of evaluating total energies for thousand atoms in a few seconds CPU time
could constitute a substantial enlargement of the domain of the 
applications of the DFT.

\begin{acknowledgments}
We acknowledge financial support from MURST (PRIN grant no. 2004023079/004). 
Discussions with Professors B. Ginatempo and I.A. Abrikosov are also gratefully
acknowledged.
\end{acknowledgments}


\end{document}